\documentclass[preprints,article,accept,pdftex,moreauthors,galaxies]{Definitions/mdpi} 
\usepackage{aas_macros}
\usepackage{graphbox}
\firstpage{1} 
\pubvolume{1}
\issuenum{1}
\articlenumber{0}
\pubyear{2022}
\copyrightyear{2022}
\datereceived{} 
\dateaccepted{} 
\datepublished{} 

\hreflink{https://doi.org/}

\Title{Probing plasma composition with the next generation Event Horizon Telescope (ngEHT)}
\TitleCitation{Plasma-Composition}

\TitleCitation{Emami, R.; Anantua, R.; Ricarte, A.; Doeleman, S.S.;  Broderick, A.; Wong, G.; Blackburn, L.; Wielgus, M.; Narayan, R.; Tremblay, G.; Alcock, C.; Hernquist, L.; Smith, R.;  Liska, M.; Natarajan, P.; Vogelsberger, M.; Curd, B.; Kramer, J.A.; Probing plasma composition with ngEHT, }

\Author{Razieh Emami $^{1}$\orcidA{}, Richard Anantua $^{3}$, Angelo Ricarte $^{1,2}$, Sheperd S. Doeleman $^{1,2}$, Avery Broderick $^{4,5}$, George Wong$^{6,7}$, Lindy Blackburn$^{1,2}$, Maciek Wielgus$^{8}$, Ramesh Narayan $^{1,2}$, Grant Tremblay $^{1}$, Charles Alcock $^{1}$, Lars Hernquist $^{1}$, Randall  Smith$^{1}$, Matthew Liska$^{1}$, Priyamvada Natarajan $^{2,9}$, Mark Vogelsberger $^{10}$, Brandon Curd $^{1,2}$, Joana A. Kramer$^{8}$}

\address{$^{1}$ \quad  Center for Astrophysics $\vert$ Harvard \& Smithsonian, 60 Garden Street, Cambridge, MA 02138, USA \\
$^{2}$ \quad Black Hole Initiative at Harvard University, 20 Garden Street, Cambridge, MA 02138, USA \\
$^{3}$ \quad Department of Physics $\&$ Astronomy, The University of Texas at San Antonio, One UTSA Circle, San Antonio, TX 78249, USA \\
$^{4}$ \quad Perimeter Institute for Theoretical Physics, 31 Caroline Street North, Waterloo, ON, N2L 2Y5, Canada \\
$^{5}$ \quad Department of Physics and Astronomy, University of Waterloo, 200 University Avenue West, Waterloo, ON, N2L 3G1, Canada \\
$^{6}$ \quad  School of Natural Sciences, Institute for Advanced Study, 1 Einstein Drive, Princeton, NJ 08540, USA \\
$^{7}$ \quad  Princeton Gravity Initiative, Princeton University, Princeton, New Jersey 08544, USA \\
$^{8}$ \quad Max-Planck-Institut f\"ur Radioastronomie, Auf dem H\"ugel 69, D-53121 Bonn, Germany \\ 
$^{9}$ \quad \quad Departments of Astronomy \& Physics, Yale University, New Haven, CT 06511, USA \\
$^{10}$ Department of Physics, Kavli Institute for Astrophysics and Space Research, Massachusetts Institute of Technology, Cambridge, MA 02139, USA
}

\corres{razieh.emami$_{-}$meibody@cfa.harvard.edu}

\abstract{We explore the plasma matter content in the innermost accretion disk/jet in M87* as relevant for an enthusiastic search for the signatures of anti-matter in the next generation of the Event Horizon Telescope (ngEHT). We model the impact of non-zero positron-to-electron ratio using different emission models including a constant electron to magnetic pressure (constant $\beta_e$ model) with a population of non-thermal electrons as well as a R-beta model populated with thermal electrons. In the former case, we pick a semi-analytic fit to the force-free region of a general relativistic magnetohydrodynamic (GRMHD) simulation, while in the latter case, we analyze the GRMHD simulations directly. In both cases, positrons are being added at the post-processing level. We generate polarized images and spectra for some of these models and find out that at the radio frequencies, both of the linear and the circular polarizations get enhanced per adding pairs. On the contrary, we show that at higher frequencies a substantial positron fraction washes out the circular polarization. We report strong degeneracies between different emission models and the positron fraction, though our non-thermal models show more sensitivities to the pair fraction than the thermal models. We conclude that a large theoretical image library is indeed required to fully understand the trends probed in this study, and to place them in the context of large set of parameters which also affect polarimetric images, such as magnetic field strength, black hole spin, and detailed aspects of the electron temperature and the distribution function.}


\begin{document}

\section{Plasma composition: Observational Studies} \label{sec:intro}
Large scale jets are launched and are collimated by their surrounding accretion flow through extracting the black hole rotational energy by purely electromagnetic mechanisms \cite{1977MNRAS.179..433B, 2010ApJ...711...50T, 2011MNRAS.418L..79T, 2012MNRAS.423L..55T,2015ASSL..414...45T}.
At radio frequencies, jets are visible from the sub-pc to the kpc scales through the synchrotron emission from the ultra-relativistic electrons gyrating inside the magnetic field. Consequently, jet links the accretion physics to the particle acceleration \cite{1979ApJ...232...34B, 1982MNRAS.199..883B}. On the theoretical front, significant progress have been made most recently in modeling the jet formation and its morphology where the current GRMHD simulations are able to recover the observed Lorentz factor. There are many factors that contribute to the jet formation, collimation, acceleration and propagation. Of the key importance is the matter content of the jet; either in terms of the normal plasma (composed of electron-protons) or the pair plasma (composed of electron-positron pairs), they lead to very distinct observational signatures. Radio-load quasars and the active galactic nuclei (AGN) make roughly 10\% of the population of jets and exhibit very powerful jets propagating hundreds of kpc away before getting disrupted \cite{1996MNRAS.283..873R}. Consequently, they can be good candidates for searching the matter content of the jet. 

Giant elliptical galaxy M87 at the center of the Virgo cluster contains a very spectacular extragalactic jet (firstly discovered by \cite{1918PLicO..13...55C}), with a relatively low radio luminosity. The synchrotron and the inverse Compton emissions from the jet bases produce radiation at other frequencies including the optical, X-rays and the $\gamma$-rays far outside the host galaxy. Consequently, M87 has been subject to wide studies at multi-frequencies from the radio to $\gamma$-rays \cite{2012ApJ...746..151A}, as well as the $X$-rays \cite{1991AJ....101.1632B}. Furthermore, there have been also an extended study on its jet collimation using the Very Long Baseline Interferometry (VLBI) at sub-millimeter wavelengths \cite{1967Natur.213..789P,1982ApJ...263..615R, 2007ApJ...668L..27K,2014ApJ...783L..33K,2015ApJ...803...30K, 2016A&A...595A..54M,2018ApJ...855..128W,2018A&A...616A.188K,2019MNRAS.486.2873C}. The synchrotron spectrum of M87 jet was first studied in \cite{1996A&A...307...61M}. The combination of all of these studies makes M87 a very valuable source to probe the matter content of the jet. 

\citet{1996MNRAS.283..873R} used the historical data from the VLBI observations of M87 at 5 GHz and probed the physical properties of the jet as well as its matter content using the synchrotron self-absorption theory \cite{1979ApJ...232...34B} relevant for the radio emission from the compact core of M87. They put constraints on the magnetic field and the particle density of the jet and eventually on the matter content of the jet. 
Their results strongly favored a pair dominated plasma, though not yet ruling out the possibility of a normal plasma. As they argued, a multi-frequency analysis of the jet may unambiguously put constraints on the matter content of the jet. 

The quasar 3C279 is another example of a luminous object in the sky, located at redshift $z=0.538$, already observed with the EHT \citep{2020A&A...640A..69K}. It is luminous from radio to $\gamma$-ray wavelengths. At the radio frequency, the VLBI observations demonstrate a very bright and unresolved core along with a jet extended to the kpc scales. There are superluminal motions associated with this source in the jet, with velocities raging from 4-15 times the speed of light, indicating a relativistic bulk speed in the jet. Consequently, the emitted radiation from the jet at different wavelengths are boosted by the Doppler effect. Owing to its relatively high flux, 3C279 is an ideal source to probe the physics of the extragalactic source. To probe the jet decomposition in quasar 3C279, \cite{1998Natur.395..457W} used the circular polarization from the observations of 3C279 at 15 GHz. 
The circular polarization is getting produced through the Faraday conversion requiring the energy distribution of particles being extended to lower energies. Combining their final results with other extragalactic sources, such as M87, they concluded that in general the extragalactic jets might be primarily composed of pairs. Furthermore, they argued that since the jet densities should be rather low, the pair dominated jet points us to a picture in which photon cascades or the Pion decay are the main origin of the radiating particles in the jet. 

The quasar 3C345 at redshift $z=0.594$ is a core dominated radio source with a prominent pc-scale jet, emitting X-rays through the Synchrotron self-Compton (SSC) process. This source has been monitored at 5 GHz every year since 1977. Higher frequency monitoring of this source, at both 10.5 and 22 GHz, are done more frequently. \cite{2000ApJ...545..100H} combined the constraints on the electron number density from the Synchrotron Self-Absorption with the kinetic luminosity constraints and concluded that C2, C3, C4, C5 and C7 components of quasar 3C345 are predominantly made of pairs than normal plasma.

Despite the fact that all of the aforementioned studies give us a positive signal on a pair plasma, others get failed in searching for positrons. 

The radio galaxy 3C 120, located at $z=0.033$, presents a strong interaction between the jet and its interstellar cloud in which the matter content of the jet are mixed with the dense surrounding thermal gas. If the jet in this source is mostly composed of pairs, their positrons would have entered to the cloud and got thermalized through the ionized energy lost. The annihilation rate of such positrons in the cloud would then have be proportional to the positron density in the jet. The observations of narrow emission lines would then inform us about the matter content of the jet. \cite{2007ApJ...665..232M} made an exploration of the matter content of the jet in this source through their annihilation line flux, using the hard X-ray and soft $\gamma$-ray spectrum from the SPI spectrometer on INTEGRAL. Their spectral analysis failed to detect any lines and thus could not consistently constrain the positron to electron ration in this source.

More statistically, \cite{1993MNRAS.264..228C} used a sample of radio-load quasars and addressed the puzzle of plasma matter content. Out of the possibility of a normal or a pair plasma and based on the annihilation constraints, combined with the assumption that pairs are originated from the inner part of the accretion, they favored an electron-proton plasma.  

In summary, there are some controversial results in the observational searches for positrons where most studies seem to favor (almost) significant fraction of pairs being presented in the plasma, while others get failed to put constraints on plasma matter content explicitly. 

Being mindful of the current status of searches for positrons, in what follows, we review different theoretical studies that motivate a pair plasma. We will then focus on two sets of toy models, one based on a semi-analytic type of models, while the other from the GRMHD simulations. In the latter case, we make the polarized images for one snapshot. Work is in progress to extend this analysis to the case with a time-averaged polarized image set for GRMHD simulations. In both cases, we add positrons as a post-processing level using different ray-tracing codes.

Taking into account the most recent EHT constraints we show that in the semi-analytical toy model, the polarimetry leads to severe constraints on the positron fraction.

%%%%%%%%%%%%%%%%%%%%%%%%%%%%%%%%%%%%
\begin{figure*}[th!]
\centering
\includegraphics[width=\textwidth]{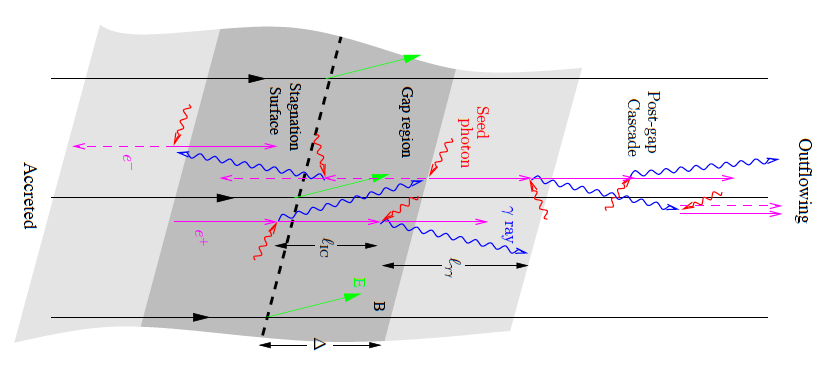}
\caption{small portion of the stagnation surface and the gap region \cite{2015ApJ...809...97B}. Inside the gap, particles are highly accelerated with a net charge separation. The scattered photons produce some pairs leading to pair catastrophe. Depending on the orientation, the created particles either get accreted to the central BH or rather populate the BH jet. 
}
\label{stagnation}
\end{figure*}
%%%%%%%%%%%%%%%%%%%%%%%%%%%%%%%%%%%%%% 

\section{Theoretical approaches in creating pairs}
There are two distinct class of models which lead to pair creation through  photon annihilation, Briett-Wheeler process \cite{1934PhRv...46.1087B}. The electron-positron pairs could either be created from a coherent, steady state and large scale mechanisms (as relevant in the gap systems) \cite{1992SvA....36..642B, 1998ApJ...497..563H,2018PhRvD..98f3016F,2018A&A...616A.184L,2018ApJ...863L..31C,2019PhRvL.122c5101P} or rather get originated from an incoherent, transient and small-scale approach (being appropriate in pair drizzle systems) \cite{2011ApJ...735....9M}. However, while in the gap approach, the high energy photons have an energy orders of magnitude above the rest-mass energy of electrons ($\simeq$ MeV), in pair drizzle approach the host photons have an energy of roughly $\mathrm{MeV}$. Despite the distinct features of the gap and drizzle models, they could be thought of as the continuum distribution in two different ends for the energy spectrum of the created photons.

In the following, we review these scenarios in more depth:

1. Gap models make high-energy photons in coherent regions with $\mathbf{E} \cdot \mathbf{B} \neq 0 $ which accelerate the leptons and make the pair cascades. 
\citet{2015ApJ...809...97B} showed that the stagnation surface, in these gap models, defined as the boundary between the material falling back to the black hole and outflows (forming the jet) would be a natural site for the pair creation followed by the particle acceleration. They showed that un-screened electric fields lead to the production of photons from the accretion flow, inside the jet. These photons would then make non-thermal particles through their inverse Compton scattering. As they argued, this method leads to a population of non-thermal particles consistent with the most recent sub mm-VLBI observations of M87. Figure \ref{stagnation} presents a small portion of stagnation surface as well as the gap region. In this method, there are two distinct mechanisms in making/propagating  the pairs. In the gap region, particles are getting highly accelerated, limited by the inverse Compton cooling, followed by the net charge separation. They emit photons which then make electron-positron pairs through the Compton scattering off the ambient soft photons. In this picture, particles moving inward are getting accreted by the central BH while those travelling outward are making the jets. In such spark gaps, positron densities may exceed the Goldreich-Julian value required to screen electric fields, limiting the efficiency of pair cascades once enough positrons are formed. 

\begin{figure*}[th!]
\centering
\includegraphics[width=0.8\textwidth]{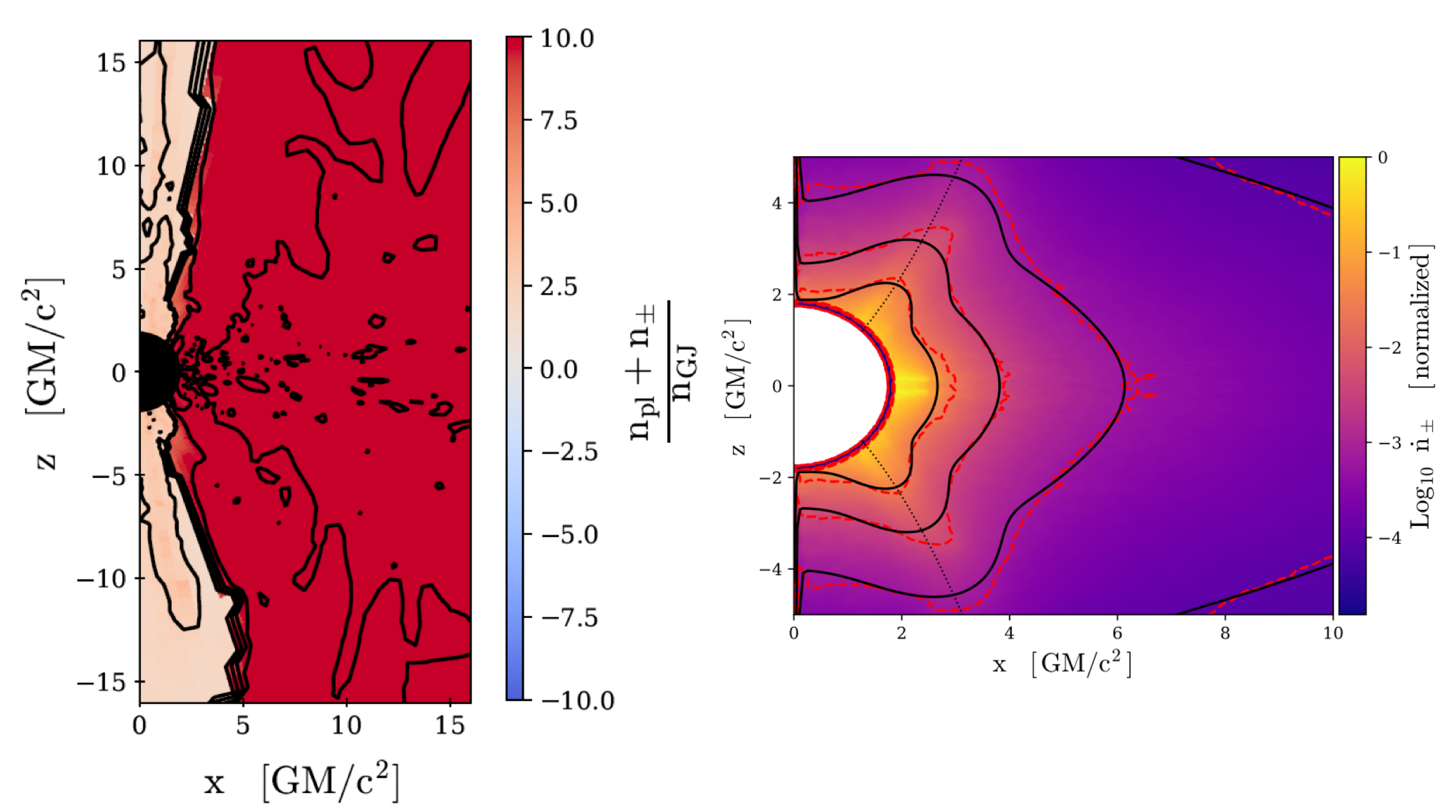}
\caption{ The time averaged pair production rate density taken from \cite{2021ApJ...907...73W} as a function of the position. The horizontal axis presents the radial coordinate while the vertical axis shows the height above the disk mid-plane.
}
\label{fig:Pair_Density}
\end{figure*}

2. Pair drizzle models predict a steady and smooth background population of photons, being created from the high-energy part of electron distribution function in the near horizon plasma. Such $\sim \mathrm{MeV}$ photons make e$^{-}$/e$^{+}$ pairs throughout their interactions. \citet{2011ApJ...735....9M} estimated the pair creation rate based on the non-relativistic GRMHD simulations. \citet{2021ApJ...907...73W} extended this approach to include a radiative based GRMHD simulation from {\tt ebh-light} code \cite{2015ApJ...807...31R, 2017ApJ...844L..24R, 2018ApJ...864..126R, 2019ascl.soft09007R}. They used an axisymmetric model and explored different mass accretion rates corresponded to an optically thin and geometrically thick accretion flow of SANE simulation. Figure \ref{fig:Pair_Density} presents the logarithmic ratio of available pair to the Goldreich-Julian density (see Eq. (33) of \cite{2021ApJ...907...73W} for more details) (left panel) for a GRMHD model with $a = 0.5$ with the accretion rate $\dot{m}/\dot{M}_{\mathrm{Edd}} = 1.1 \times 10^{-5}$. From the plot, it is evident that 
there are substantial chance to have pairs in the disk and in the funnel jet. The right panel also shows the logarithmic rate of the pair production density in the same model. It is seen that the rate of pair production is enhanced in the inner part of the disk and in the equatorial plane. It is however diminishes significantly at the outer edge of the disk as well as the high scale heights. 

Consequently, we conclude that while in gap models pairs are being mostly created in the jet funnel, pair drizzle models make them more in the accretion disk. 

%%%%%%%%%%%%%%%%%%%%%%%%%%%%%%%%%%%%
%%%%%%%%%%%%%%%%%%%%%%%%%%%%%%%%%%%%%%

%%%%%%%%%%%%%%%%%%%%%%%%%%%%%%%%%%%%
\begin{figure*}[th!]
\centering
\includegraphics[width=\textwidth]{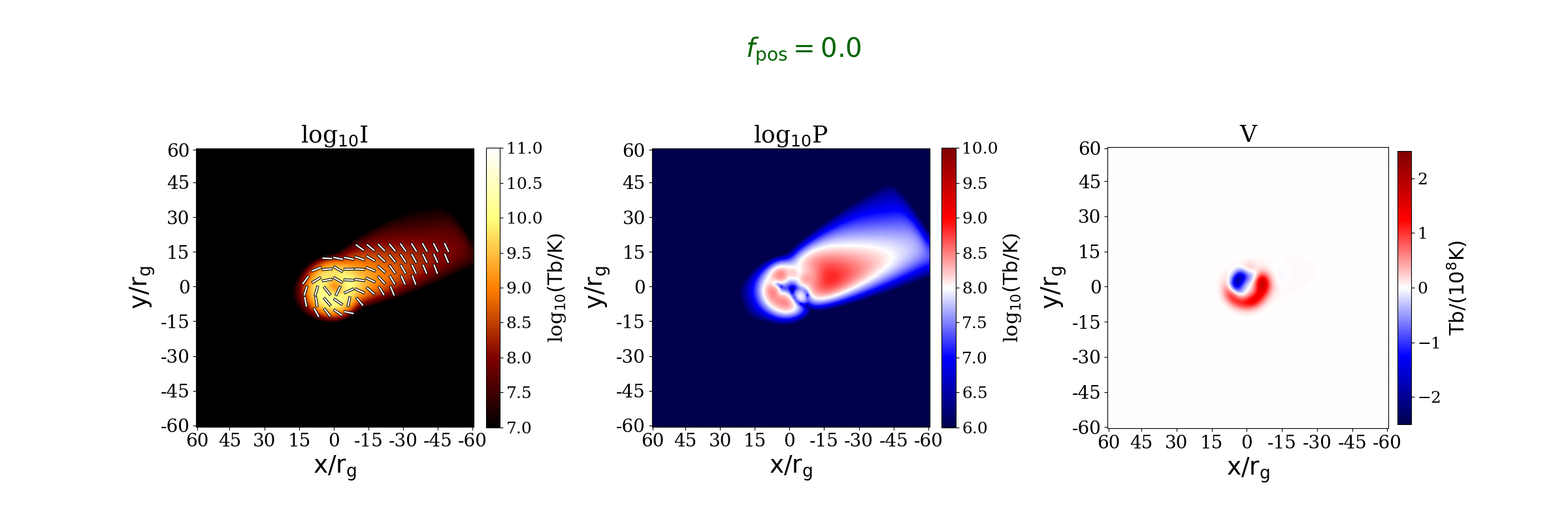}
\includegraphics[width=\textwidth]{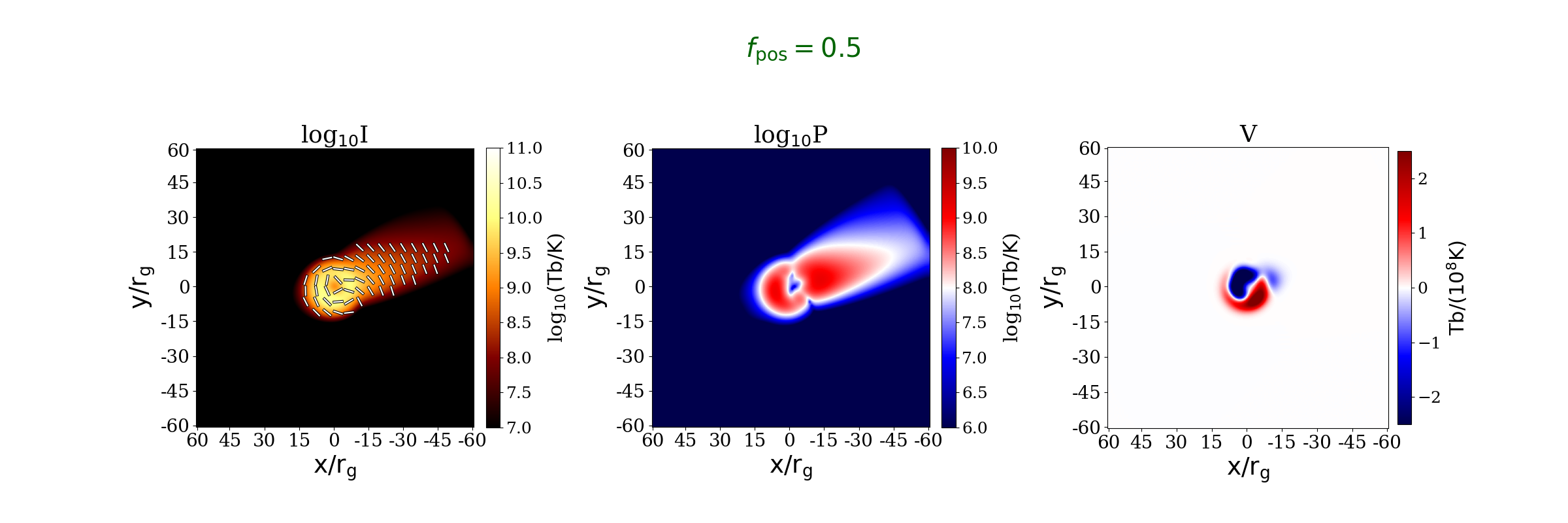}
\caption{Down rows present the polarized map of M87* inferred from the constant $\beta$ model at 230 GHz. From the top to bottom, we present the images for $f_{\mathrm{pos}} = 0.0, 0.5$, respectively.}
\label{fig:Best-Bet-image}
\end{figure*}
%%%%%%%%%%%%%%%%%%%%%%%%%%%%%%%%%%%%%%

\section{Radiative Transfer of Pair Plasma}
Due to their opposite charge, positrons gyrate in a magnetic field in the opposite direction to electrons. This has significant effects on the radiative transfer coefficients, namely the synchrotron emission/absorption and Faraday rotation. Compared to an ionic plasma, a pair plasma produces no Faraday rotation, nor does it produce any circular polarization from synchrotron radiation. On the other hand, linear polarization emission and Faraday conversion persist. Consequently, polarization can be used to place constraints on the plasma composition and has fueled a decades-long debate about the composition of astrophysical jets \citep{Blandford&Konigl1979,Wardle+1998,Wardle&Homan2003}.   

In GRRT, the radiative transfer coefficients \citep[see e.g.,][for their derivation and definitions]{Dexter2016} are modified via:
\begin{align}
    j_{I,Q,U} &\to (1+f) j_{I,Q,U}, \nonumber \\
    j_{V} &\to (1-f) j_{V}, \nonumber \\
    \alpha_{I,Q,U} &\to (1+f) \alpha_{I,Q,U}, \nonumber \\
    \alpha_{V} &\to (1-f) \alpha_{V}, \nonumber \\
    \rho_{Q,U} &\to (1+f)\rho_{Q,U}, \nonumber \\
    \rho_{V} &\to (1-f)\rho_{V}.
\end{align}
where $j_i$ and $\alpha_i$ are the emission and absorption coefficients, $\rho_i$ refer to the rotativities, and $f$ describes the positron-to-electron ratio. Qualitatively, increasing the positron fraction tends to increase the linear polarization fraction by reducing the Faraday rotation depth. The circular polarization fraction and the image morphology may also change significantly as the balance between circular polarization generated by emission and Faraday conversion changes. Since the Faraday conversion, Faraday rotation, and circular polarization emission coefficients evolve differently with the frequency, this may lead to large changes in the polarized spectrum. Consequently, a multi-frequency analysis might be very helpful in putting constraints on the plasma composition. 

Below, we use the above prescription and analyse the impact of positrons in polarized images from a semi-analytical models \ref{semi-analytic} as well as two sets of GRMHD simulation \ref{GRMHD_sim}. As we will describe, in both cases, we have added the positrons at the ray-tracing level by the aforementioned algorithm. Throughout our analysis, we merely focus on M87*.  

%%%%%%%%%%%%%%%%%%%%%%%%%%%%%%%%%%%%
\begin{figure*}[th!]
\centering
\includegraphics[width=\textwidth]{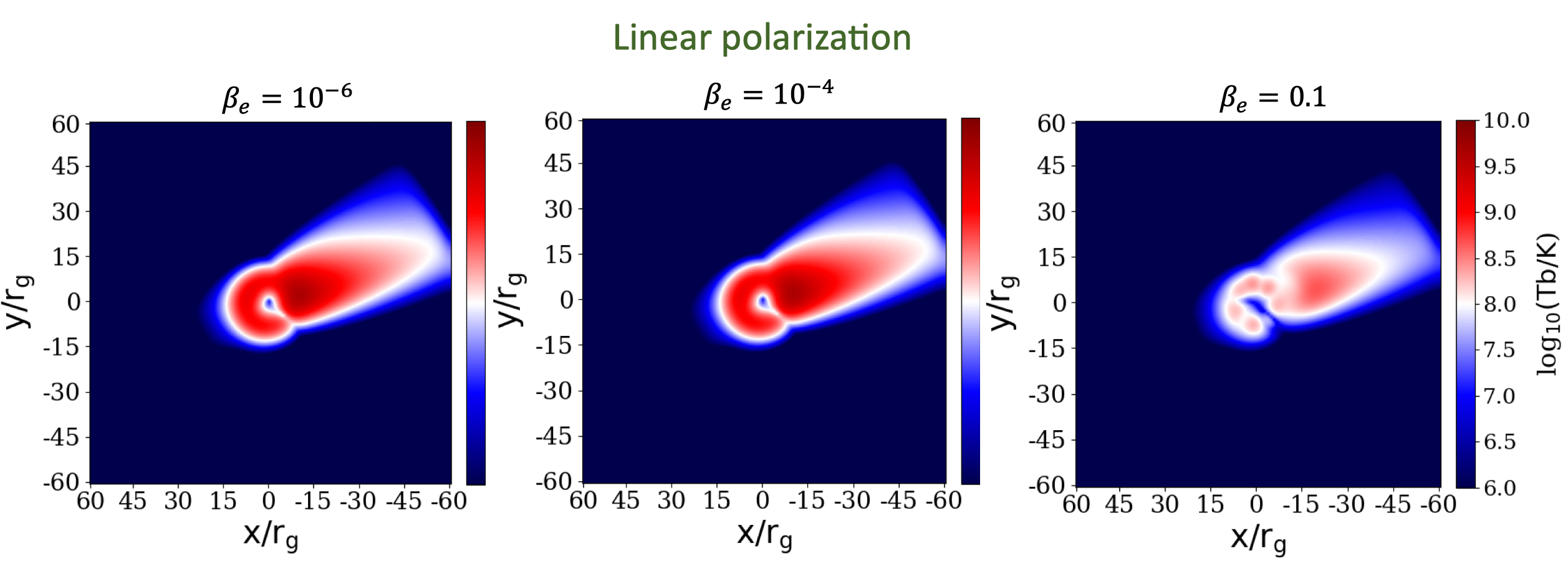}
\includegraphics[width=\textwidth]{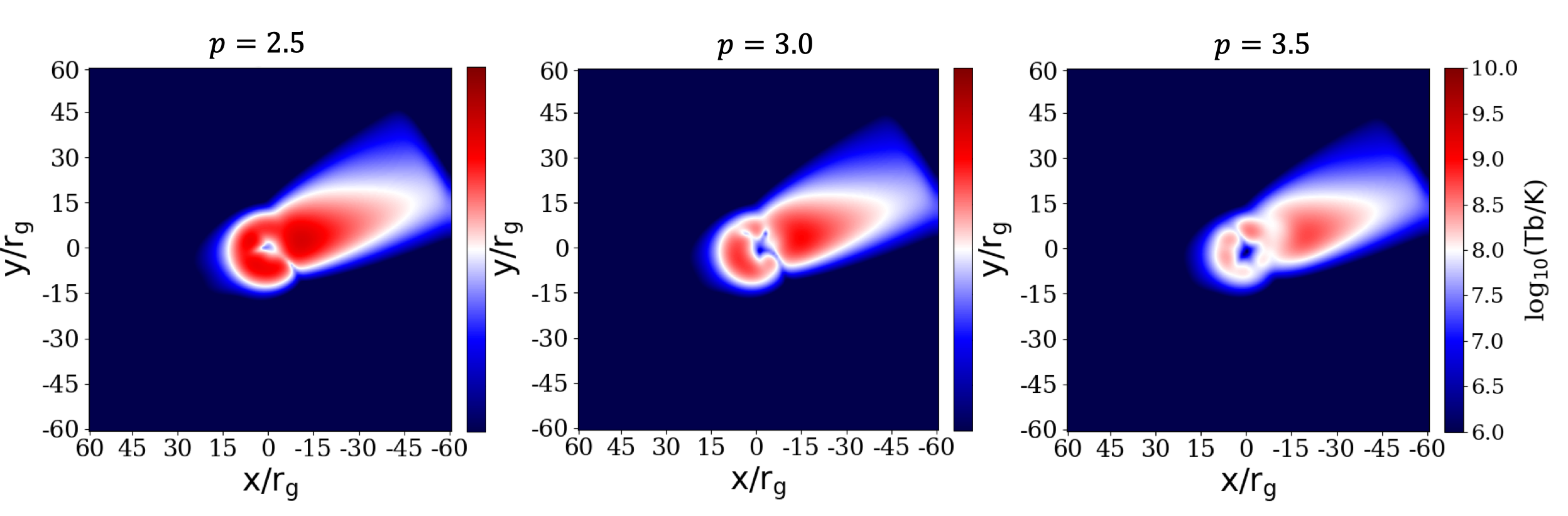}
\caption{The (log)-linear polarization map for the constant $\beta$ model with $f_{\mathrm{pos}}$=0.1. Top panel shows the case with p=3.2 and with changing $\beta_e = 10^{-6}, 10^{-4}, 10^{-1}$. Bottom panel presents the case with fixed $\beta_e = 10^{-2}$ while changing p = 2.5, 3.0, 3.5.}
\label{fig:linear-pol-models}
\end{figure*}
%%%%%%%%%%%%%%%%%%%%%%%%%%%%%%%%%%%%%%
\section{Positron effects in the semi-analytical models} \label{semi-analytic}
To probe the observational signatures of a pair plasma in the polarized emission from the jet/accretion in M87*, we make use of a semi-analytical approach \cite{2021ApJ...923..272E}, as a self-similar jet model focusing on the force-free regions of a Blandford-Znajek outflow model in \cite{Anantua2020a,Blandford2017} with several extensions including a general relativistic ray-tracing from GR-Trans code as well as the usage of non-thermal distribution as the number density of pairs. In this model, the number density of pairs is mapped to total electron/positron pressure, with an overall electron to magnetic pressure $\beta_e$. Furthermore, we assume a self-similar parabolic jet profile $\xi = s^2/z$, in the cylindrical coordinate. We use some fitting formulae for the $\xi$ using the magnetic flux $\Phi_B(\xi)$, the line angular speed $\Omega_B(\xi)$ and the velocity component along with the z-coordinate, $v_z(\xi)$ being extracted from a {\tt HARM} simulation \cite{McKinney2012}. In more detail, for the GRMHD, we use a magnetically arrested disk (MAD) simulation with $a/M$ = 0.92 and infer the fitting formulas at $z=50 M$. Note that these models are idealized and axis-symmetric, and polarized signatures are sensitive to additional parameters, including spin, magnetic field state, and the electron-to-ion temperature ratio \citep{2021ApJ...910L..13E}. Nevertheless, this simple model can be used to explore and illustrate the qualitative changes as positrons are added to the accretion flow.

%%%%%%%%%%%%%%%%%%%%%%%%%%%%%%%%%%%%
\begin{figure*}[th!]
\centering
\includegraphics[width=\textwidth]{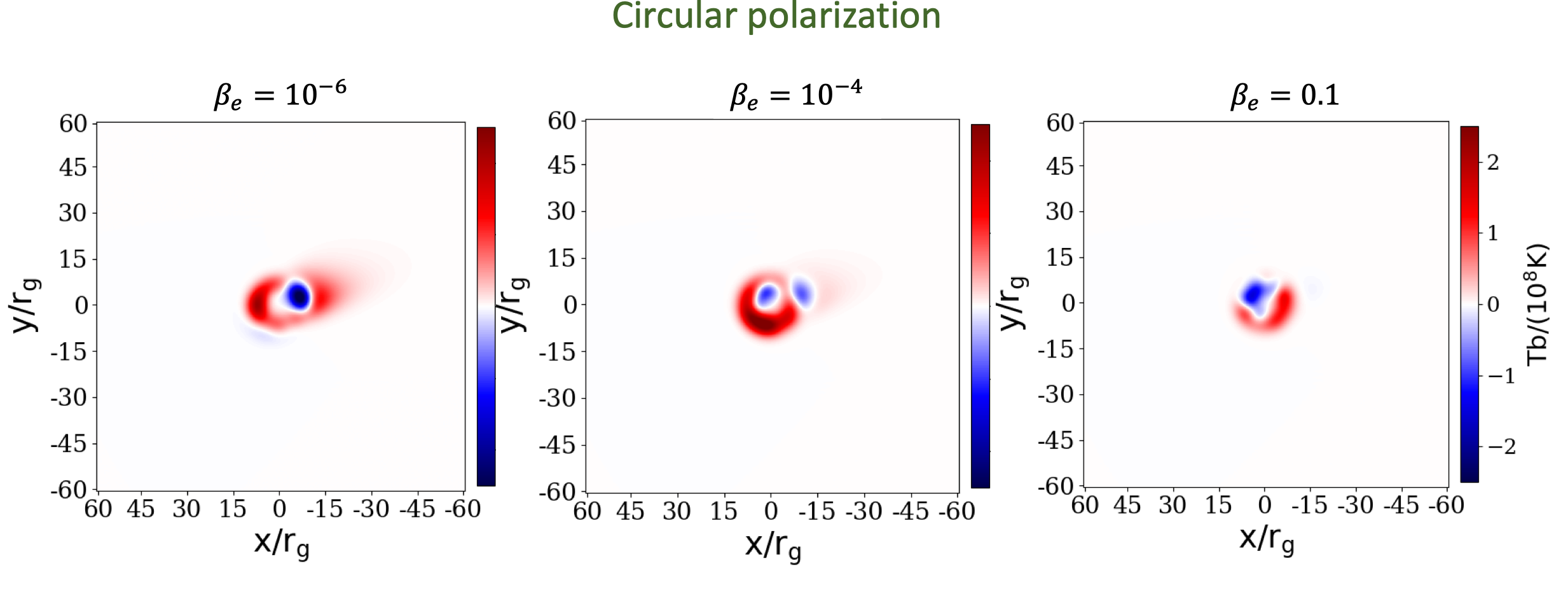}
\includegraphics[width=\textwidth]{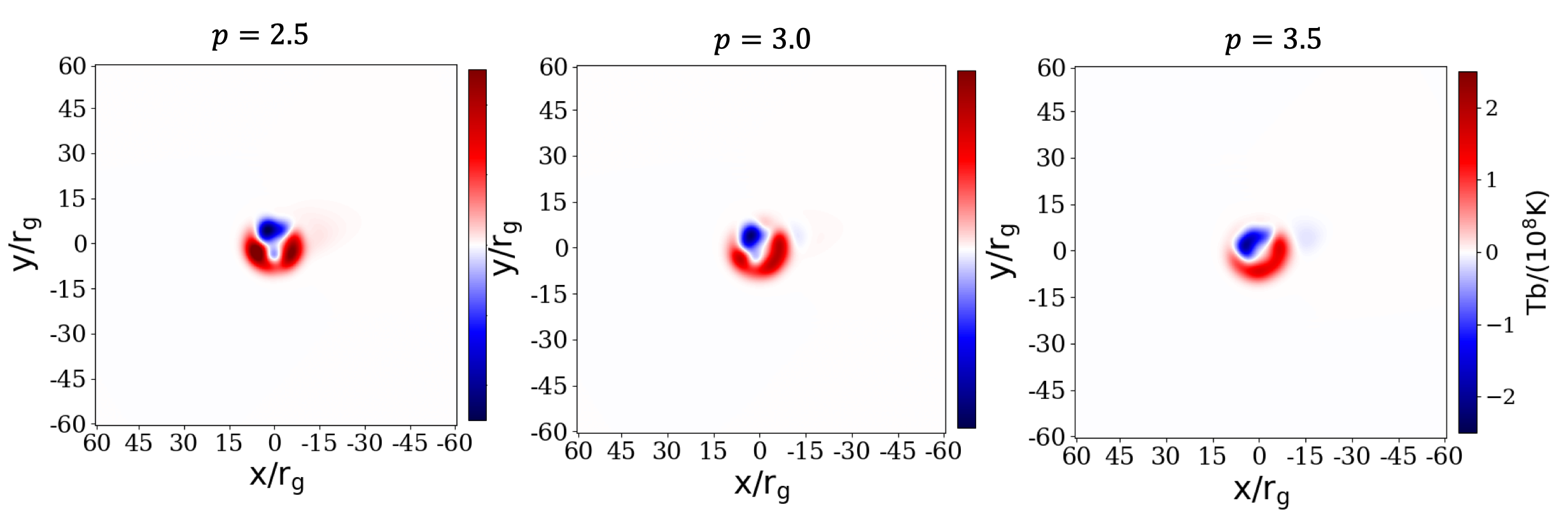}
\caption{The circular polarization map for the constant $\beta$ model with $f_{\mathrm{pos}}$=0.1. Top panel shows the case with p=3.2 and with changing $\beta_e = 10^{-6}, 10^{-4}, 10^{-1}$. Bottom panel presents the case with fixed $\beta_e = 10^{-2}$ while changing p = (2.5, 3.0, 3.5).}
\label{fig:circular-pol-models}
\end{figure*}
%%%%%%%%%%%%%%%%%%%%%%%%%%%%%%%%%%%%%%
Next, we extrapolate these quantities to larger radii taking advantage of the self-similarity in $\xi$; see Eqs. (18-26) of \cite{2021ApJ...923..272E} for more details. 

As already mentioned above, we use a non-thermal, power-low distribution for the emitting electrons, with a Lorentz factor between $\gamma_{\mathrm{min}}$ and $\gamma_{\mathrm{max}}$:
\begin{equation}
N_{e^-}(\gamma) = 
\begin{cases}
N_0\gamma^{-p} &\gamma_{\rm min}\leq \gamma\leq\gamma_{\rm max}\\
0 & \mathrm{otherwise}
\end{cases},
\label{eq:powerlaw}
\end{equation}
where $N_0 = n_{e^-} (p-1)/\left(\gamma_{\rm min}^{1-p}-\gamma_{\rm max}^{1-p}\right)$ refers to the overall normalization of the electron distribution and with $n_{e^{-}}$ describes the total number density of electrons as given by Eq. (28) of \cite{2021ApJ...923..272E}.  
We implement this model to GR-Trans, modifying its equations for the emissivity, absorption and the Faraday terms to include the contribution of positrons. Below, we describe the results including the polarized images, the spectra and the multi-frequency analysis.

\subsection{Polarized Images} \label{polarized-image-semi-analytic}
Figure \ref{fig:Best-Bet-image} presents the polarized image of the constant $\beta_e$ model at 230 GHz for the case with no positrons (top panel) and the one with 50\% of positrons (bottom panel). From the plot, it is seen that in this model, adding the positrons significantly increases the linear and the circular polarization. Consequently, we expect that the EHT polarimetric constraints disfavor the case with significant pair fraction in the constant $\beta_e$ model. 

Next, we explore the impact of changing the parameters of the emission model. Figures \ref{fig:linear-pol-models}-\ref{fig:circular-pol-models} show the linear and circular polarization maps for different models parameters. In each figure, in the top panel, we change $\beta_e = 10^{-6}, 10^{-4}, 10^{-2}$, while in the bottom row, we alter the slope $p = 2.5, 3.0, 3.5$, while fixing $f_{\mathrm{pos}}$ = 0.1. From the plots, it is inferred that increasing the $\beta_e$ as well as $p$ suppress the linear and circular polarization. Consequently, we may expect that models with higher values of $\beta_e$ as well as the non-thermal slope have a better chance to satisfy the current polarimetric constraints from the EHTC \cite{2021ApJ...910L..13E, 2021ApJ...923..272E}. Finally, it is seen that the shape of the circular polarization also changes when we vary the parameters of the emission model. This imply that a direct detection of the circular polarization map might be useful to break the degeneracy between the physical parameters of the model as well as the matter content of the emitting plasma.

\subsection{Spectral analysis} \label{spectral-analysis}
\begin{figure*}[th!]
\centering
\includegraphics[width=\textwidth]{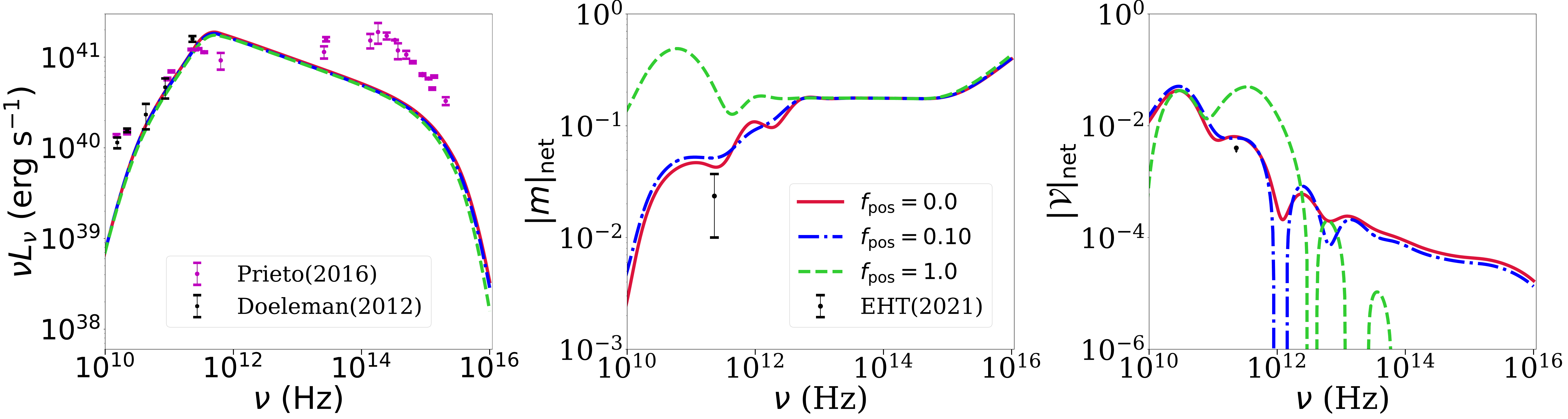}
\caption{The spectrum of the best bet model for the jet model \cite{2021ApJ...923..272E}. From the left to right, we present the total intensity, the linear and the circular polarization, respectively. Overlaid on each panel, we also present the observational data points. }
\label{fig:Best-Bet-Spectrum}
\end{figure*}

In Figure \ref{fig:Best-Bet-Spectrum} we analyze the spectrum of the total intensity, the linear and circular polarization. $\beta_e = 10^{-2}$ and the power-low index $p=3.2$ are kept fixed, while we vary the $f_{\mathrm{pos}} = (0.0, 0.1, 1.0)$. Overlaid in each panel, we present the observational data points from \cite{Doeleman2008, Doeleman2012, Prieto2016} as well as the most recent results from the EHT observation \cite{2021ApJ...910L..13E, 2021ApJ...923..272E}. Furthermore, we fix the model parameters to match the observed flux at $\nu$= 230 GHz. Consequently, the flux is off at higher frequencies. This implies that while the current toy model describes the radio observations quite well, a more complicated model is required to get the flux right at higher frequencies. The fractional linear polarization, from the middle panel, shows more sensitivity to varying the positron fraction. Based on our simple semi-analytic model, it is inferred that our models with higher $f_{\mathrm{pos}}$ sit above the current EHT constraints. In \citet{2021ApJ...923..272E}, we made a detailed survey of different models and realized that varying the $\beta_e$ and $p$ do not change the above conclusion, implying that there are severe constraints for the constant $\beta_e$ model. However, we emphasize that this conclusion may not be easily generalized to other accretion flow models. For example, as we will see below, certain thermal models provide less severe constraints. Finally, the circular polarization, in the right panel, establishes more interesting dependencies to the positron fraction. It is explicitly seen that the circular polarization is not only too sensitive to the positron fraction at the radio frequencies, but it also shows distinct features at the higher frequencies. Consequently, a multi-frequency analysis might break the degeneracies in searching for positrons. 

%%%%%%%%%%%%%%%%%%%%%%%%%%%%%%%%%%%%
%%%%%%%%%%%%%%%%%%%%%%%%%%%%%%%%%%%%%%
\subsection{Multi-frequency analysis} \label{multi-frequency}
\begin{figure*}[th!]
\centering
\includegraphics[width=\textwidth]{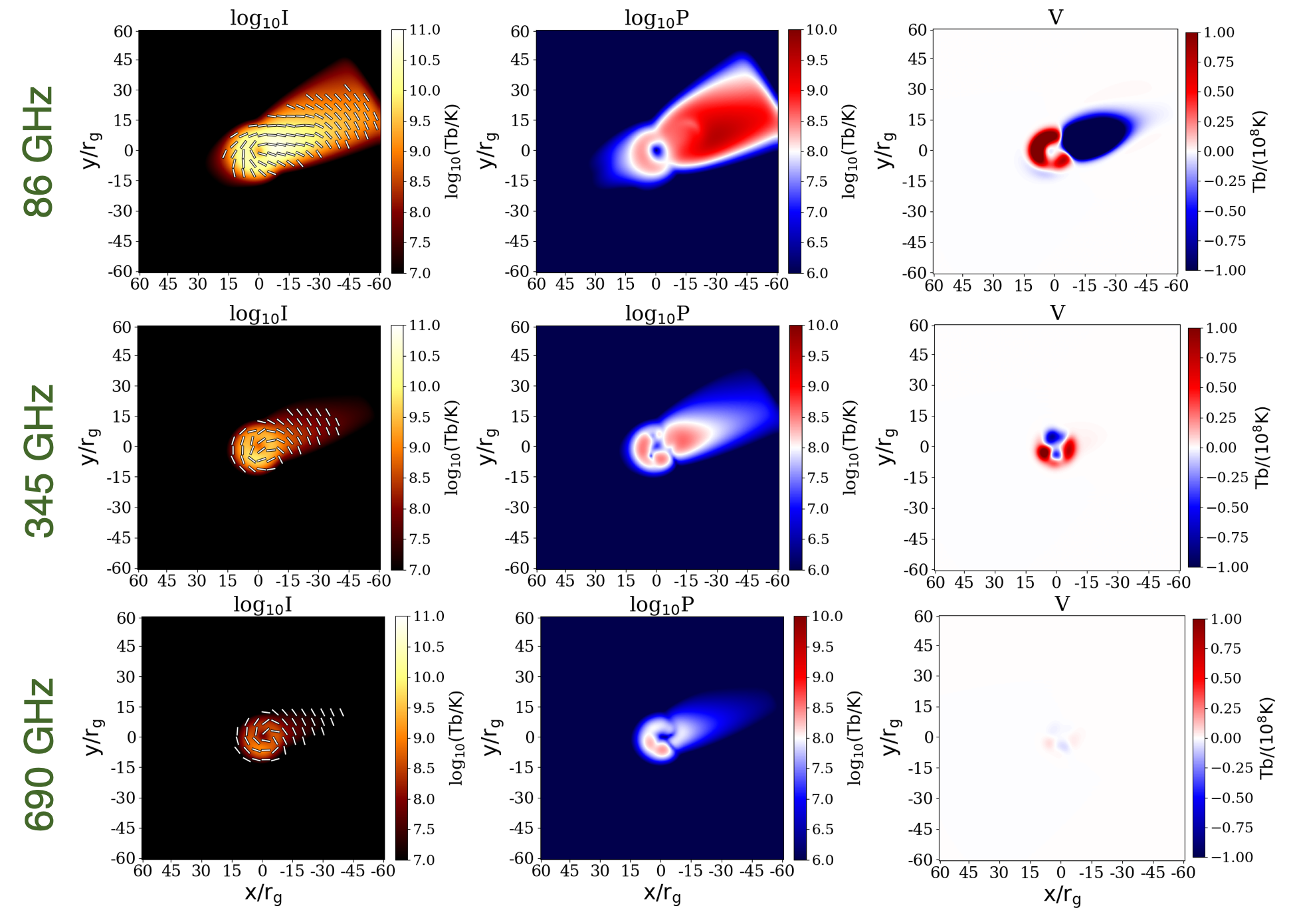}
\caption{Down rows present the polarized map (left to right the intensity, linear and circular polarization) of M87* in constant beta model at 86, 345 and 690 GHz, respectively. The plot is for f$_{\mathrm{pos}}$ = 0.1, p=3.2 and $\beta = 10^{-2}$. }
\label{fig:multi-frequency}
\end{figure*}

Figure \ref{fig:multi-frequency} presents a multi-frequency image analysis at 86, 345 and 690 GHz, as relevant for the ngEHT. In each row, from the left to right, we present the intensity, linear and the circular polarizations. From the plot, it is inferred that increasing the frequency washes out the linear and circular polarization substantially. Furthermore, the image is also core shifted at higher $\nu$s where the larger scale patterns in the EVPAs are getting boosted to the central part of the image. Finally, the shape of the circular polarization is also altered at higher frequencies. 
%%%%%%%%%%%%%%%%%%%%%%%%%%%%%%%%%%%%
%%%%%%%%%%%%%%%%%%%%%%%%%%%%%%%%%%%%%%
\section{Positrons in GRMHD KHARMA Simulation} \label{GRMHD_sim}
Next, we generalize the above semi-analytic approach to a direct GRMHD simulation approach (Anantua et al.~in prep.), with the positrons are still being added during a post-processing step to KHARMA simulation \cite{2021JOSS....6.3336P}. While an in depth analysis of the impact of positrons is left to a separate work, Anantua et al.~in prep., here we aim to take a first look at the possible importance of changing the accretion type as well as changing electron emission profiles from non-thermal models to thermal models.  

\begin{figure*}[th!]
\centering
\includegraphics[width=\textwidth]{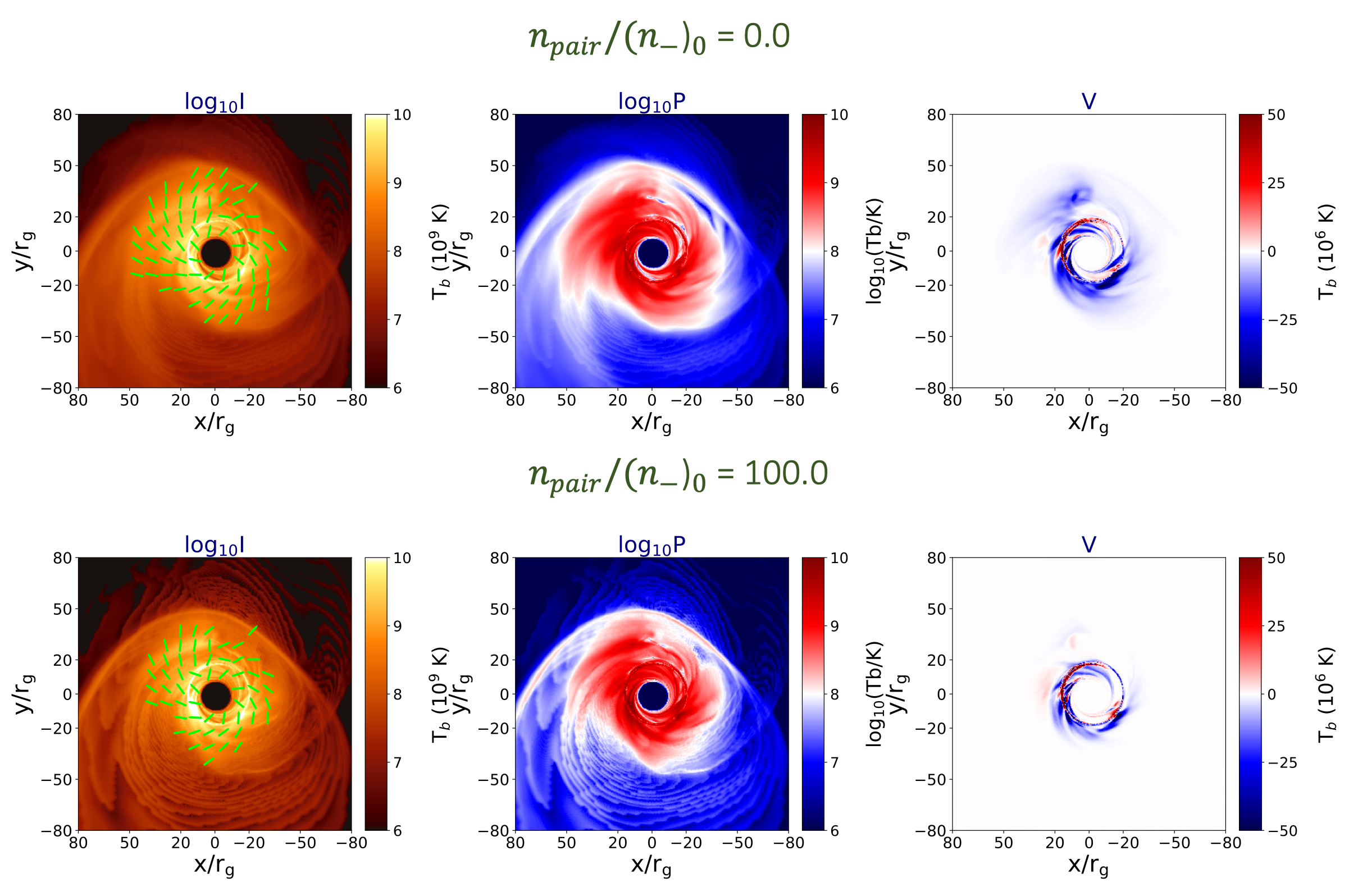}
\caption{A map from MAD simulations with a = +0.94 at $R_{\mathrm{high}}$ = 20 in R-beta model from KHARMA simulations. The top row refers to the case with no positrons while the bottom one describes the case with 100\% positrons compared with the original number of electrons.}
\label{fig:MAD_Kharma_Rhigh20_Rbeta}
\end{figure*}

\begin{figure*}[th!]
\centering
\includegraphics[width=\textwidth]{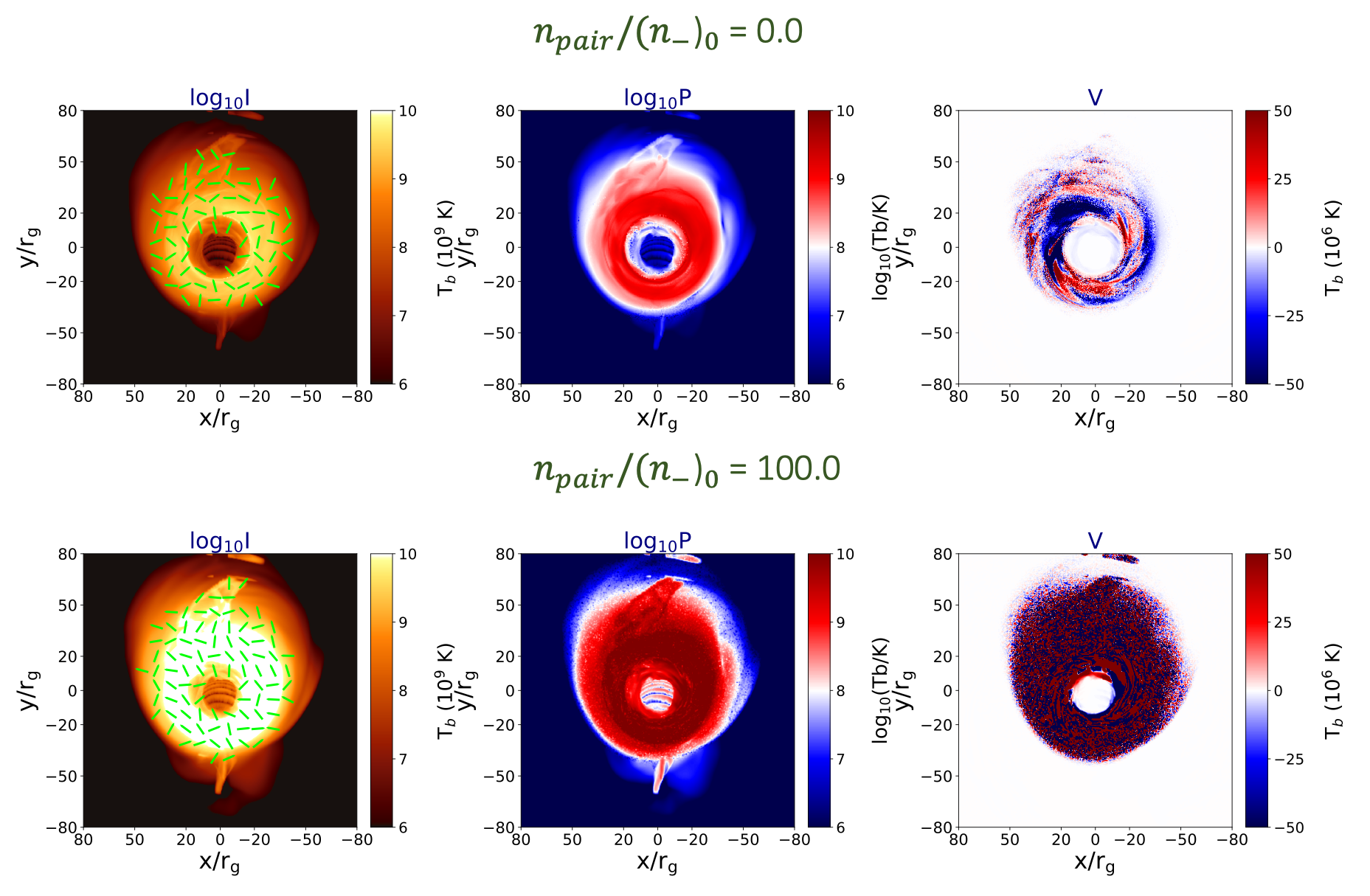}
\caption{A map from SANE simulations with a = +0.5 at $R_{\mathrm{high}}$ = 20 in R-beta model from KHARMA simulations. The top row refers to the case with no positrons while the bottom one describes the case with 100\% positrons compared with the original number of electrons.}
\label{fig:SANE_Kharma_Rhigh20_Rbeta}
\end{figure*}

In Figs. \ref{fig:MAD_Kharma_Rhigh20_Rbeta} and \ref{fig:SANE_Kharma_Rhigh20_Rbeta}, we present a MAD model with $a=0.94$ as well as a SANE model with $a=+0.5$, respectively,  each with $R_\mathrm{high}=20$. In the top row, we show the result of a normal ray tracing, without the addition of pairs, using {\sc ipole} \citep{Moscibrodzka&Gammie2018}, while in the bottom one, we present the case with 100 times more electron-positron pairs, added at a post-processing level, to the pre-existing electron number density in KHARMA GRMHD simulation. This approximates a fully pair plasma while preserving charge neutrality. From left to right, different columns present the total intensity,  linear and the  circular polarization, respectively. 

In Figure \ref{fig:MAD_Kharma_Rhigh20_Rbeta}, the addition of positrons has remarkably little effect on the linear polarization pattern. A more significant difference is seen in circular polarization, however, which historically has been promising to use to test plasma composition models \citep{Wardle+1998}. This is because only ionic plasma can produce the circular polarization via direct synchrotron emission, and a pair plasma does not perform Faraday rotation, which can affect linear polarization which goes through Faraday conversion \citep{Wardle&Homan2003}. Since large scale circularly polarized emission originates from direct synchrotron in this model \citep{2021MNRAS.505..523R}, the large scale emission disappears when a significant population of pairs is added.

%%%%%%%%%%%%%%%%%%%%%%%%%%%%%%%%%%%%
%%%%%%%%%%%%%%%%%%%%%%%%%%%%%%%%%%%%%%
In Figure \ref{fig:SANE_Kharma_Rhigh20_Rbeta}, we see that the linear polarization pattern of the SANE model is much more strongly affected by the addition of pairs than in the MAD model. This is due because the model is intrinsically Faraday thick, resulting in a significant scrambling. The addition of pairs dramatically decreases the Faraday depth, resulting in an ordered linear polarization pattern. Faraday rotation may also have an indirect effect on circular polarization, by scrambling the linear polarization that would be transformed into circular polarization. As we see here, the addition of pairs dramatically increases the resolved circular polarization by removing this scrambling effect.

Comparing the current GRMHD simulation results with that of Sec. \ref{semi-analytic}, it is inferred that changing the emission model may significantly affect the morphology of the polarized images. 

%%%%%%%%%%%%%%%%%%%%%%%%%%%%%%%%%%%%%%%%
%%%%%%%%%%%%%%%%%%%%%%%%%%%%%%%%%%%%%%%

\section{Conclusion} \label{coclusion}
There have been many outstanding observational claims for a substantially pair dominated plasma in the radio-load quasars and in AGNs. Motivated by this, in this manuscript, we studied the theoretical signatures of having non-zero positrons in the polarized images in a force free, semi-analytic jet model (Sec. \ref{semi-analytic}) as well as a snapshot of the KHARMA GRMHD simulations (Sec. \ref{GRMHD_sim}), where in both cases, positrons are added at the post-processing level. While the Faraday rotation diminishes by increasing the positron fraction, the Faraday Conversion is boosted linearly. We showed that the role of positrons is also mixed with different emission models in a degenerate picture. A multi-frequency analysis in the radio band, 86-690 GHz as relevant for the ngEHT, or a much wider frequency search, in the near infrared and the X-ray, may however break the degeneracy. Increasing the frequency, there is a core shift in images which washes out the power at larger scales and squeeze them more to the central part of the image. Consequently, we argue that lower frequencies may give rise to a better chance to search for positrons. Therefore, 86 or 230 GHz from the ngEHT could be a very relevant frequencies to look for a pair plasma in the heart of M87*. Finally, we emphasis that the positron fraction is just one of many physical parameters that affect the polarized images. In a future work, we aim for a broader investigation including a large image library probing different GRMHD models and emission prescriptions to determining how well these results may be generalized.

\section{Acknowledgements}
We greatly appreciate positrons for most likely making the AGN jets and we wish to discover them using the ngEHT! Razieh Emami acknowledges the support by the Institute for Theory and Computation at the Center for Astrophysics as well as grant numbers 21-atp21-0077, NSF AST-1816420 and HST-GO-16173.001-A for very generous supports. This work was supported in part by the Black Hole Initiative, which is supported by grants from the Gordon and Betty Moore Foundation and the John Templeton Foundation. The opinions expressed in this publication are those of the author(s) and do not necessarily reflect the views of the Moore or Templeton Foundations.

\section{References}
\bibliography{ms}

\begin{thebibliography}{999}

\bibitem[{Blandford} and {Znajek}(1977)]{1977MNRAS.179..433B}
{Blandford}, R.D.; {Znajek}, R.L.
\newblock {Electromagnetic extraction of energy from Kerr black holes.}
\newblock {\em \mnras} {\bf 1977}, {\em 179},~433--456.
\newblock {\url{https://doi.org/10.1093/mnras/179.3.433}}.

\bibitem[{Tchekhovskoy} \em{et~al.}(2010){Tchekhovskoy}, {Narayan}, and
  {McKinney}]{2010ApJ...711...50T}
{Tchekhovskoy}, A.; {Narayan}, R.; {McKinney}, J.C.
\newblock {Black Hole Spin and The Radio Loud/Quiet Dichotomy of Active
  Galactic Nuclei}.
\newblock {\em \apj} {\bf 2010}, {\em 711},~50--63,
  \href{http://xxx.lanl.gov/abs/0911.2228}{{\normalfont
  [arXiv:astro-ph.HE/0911.2228]}}.
\newblock {\url{https://doi.org/10.1088/0004-637X/711/1/50}}.

\bibitem[{Tchekhovskoy} \em{et~al.}(2011){Tchekhovskoy}, {Narayan}, and
  {McKinney}]{2011MNRAS.418L..79T}
{Tchekhovskoy}, A.; {Narayan}, R.; {McKinney}, J.C.
\newblock {Efficient generation of jets from magnetically arrested accretion on
  a rapidly spinning black hole}.
\newblock {\em \mnras} {\bf 2011}, {\em 418},~L79--L83,
  \href{http://xxx.lanl.gov/abs/1108.0412}{{\normalfont
  [arXiv:astro-ph.HE/1108.0412]}}.
\newblock {\url{https://doi.org/10.1111/j.1745-3933.2011.01147.x}}.

\bibitem[{Tchekhovskoy} and {McKinney}(2012)]{2012MNRAS.423L..55T}
{Tchekhovskoy}, A.; {McKinney}, J.C.
\newblock {Prograde and retrograde black holes: whose jet is more powerful?}
\newblock {\em \mnras} {\bf 2012}, {\em 423},~L55--L59,
  \href{http://xxx.lanl.gov/abs/1201.4385}{{\normalfont
  [arXiv:astro-ph.HE/1201.4385]}}.
\newblock {\url{https://doi.org/10.1111/j.1745-3933.2012.01256.x}}.

\bibitem[{Tchekhovskoy}(2015)]{2015ASSL..414...45T}
{Tchekhovskoy}, A.
\newblock {Launching of Active Galactic Nuclei Jets}.
\newblock In Proceedings of the The Formation and Disruption of Black Hole
  Jets; {Contopoulos}, I.; {Gabuzda}, D.; {Kylafis}, N., Eds.,  2015, Vol. 414,
  {\em Astrophysics and Space Science Library}, p.~45.
\newblock {\url{https://doi.org/10.1007/978-3-319-10356-3_3}}.

\bibitem[{Blandford} and {K{\"o}nigl}(1979)]{1979ApJ...232...34B}
{Blandford}, R.D.; {K{\"o}nigl}, A.
\newblock {Relativistic jets as compact radio sources.}
\newblock {\em \apj} {\bf 1979}, {\em 232},~34--48.
\newblock {\url{https://doi.org/10.1086/157262}}.

\bibitem[{Blandford} and {Payne}(1982)]{1982MNRAS.199..883B}
{Blandford}, R.D.; {Payne}, D.G.
\newblock {Hydromagnetic flows from accretion disks and the production of radio
  jets.}
\newblock {\em \mnras} {\bf 1982}, {\em 199},~883--903.
\newblock {\url{https://doi.org/10.1093/mnras/199.4.883}}.

\bibitem[{Reynolds} \em{et~al.}(1996){Reynolds}, {Fabian}, {Celotti}, and
  {Rees}]{1996MNRAS.283..873R}
{Reynolds}, C.S.; {Fabian}, A.C.; {Celotti}, A.; {Rees}, M.J.
\newblock {The matter content of the jet in M87: evidence for an
  electron-positron jet}.
\newblock {\em \mnras} {\bf 1996}, {\em 283},~873--880,
  \href{http://xxx.lanl.gov/abs/astro-ph/9603140}{{\normalfont
  [arXiv:astro-ph/astro-ph/9603140]}}.
\newblock {\url{https://doi.org/10.1093/mnras/283.3.873}}.

\bibitem[{Curtis}(1918)]{1918PLicO..13...55C}
{Curtis}, H.D.
\newblock {The Planetary Nebulae}.
\newblock {\em Publications of Lick Observatory} {\bf 1918}, {\em 13},~55--74.

\bibitem[{Abramowski} \em{et~al.}(2012){Abramowski}, {Acero}, {Aharonian},
  {Akhperjanian}, {Anton}, {Balzer}, {Barnacka}, {Barres de Almeida},
  {Becherini}, {Becker}, {Behera}, {Bernl{\"o}hr}, {Birsin}, {Biteau},
  {Bochow}, {Boisson}, {Bolmont}, {Bordas}, {Brucker}, {Brun}, {Brun}, {Bulik},
  {B{\"u}sching}, {Carrigan}, {Casanova}, {Cerruti}, {Chadwick}, {Charbonnier},
  {Chaves}, {Cheesebrough}, {Clapson}, {Coignet}, {Cologna}, {Conrad},
  {Dalton}, {Daniel}, {Davids}, {Degrange}, {Deil}, {Dickinson},
  {Djannati-Ata{\"\i}}, {Domainko}, {Drury}, {Dubus}, {Dutson}, {Dyks},
  {Dyrda}, {Egberts}, {Eger}, {Espigat}, {Fallon}, {Farnier}, {Fegan},
  {Feinstein}, {Fernandes}, {Fiasson}, {Fontaine}, {F{\"o}rster},
  {F{\"u}{\ss}ling}, {Gallant}, {Gast}, {G{\'e}rard}, {Gerbig}, {Giebels},
  {Glicenstein}, {Gl{\"u}ck}, {Goret}, {G{\"o}ring}, {H{\"a}ffner}, {Hague},
  {Hampf}, {Hauser}, {Heinz}, {Heinzelmann}, {Henri}, {Hermann}, {Hinton},
  {Hoffmann}, {Hofmann}, {Hofverberg}, {Holler}, {Horns}, {Jacholkowska}, {de
  Jager}, {Jahn}, {Jamrozy}, {Jung}, {Kastendieck}, {Katarzy{\'n}ski}, {Katz},
  {Kaufmann}, {Keogh}, {Khangulyan}, {Kh{\'e}lifi}, {Klochkov}, {Klu{\'z}niak},
  {Kneiske}, {Komin}, {Kosack}, {Kossakowski}, {Laffon}, {Lamanna}, {Lennarz},
  {Lohse}, {Lopatin}, {Lu}, {Marandon}, {Marcowith}, {Masbou}, {Maurin},
  {Maxted}, {Mayer}, {McComb}, {Medina}, {M{\'e}hault}, {Moderski}, {Moulin},
  {Naumann}, {Naumann-Godo}, {de Naurois}, {Nedbal}, {Nekrassov}, {Nguyen},
  {Nicholas}, {Niemiec}, {Nolan}, {Ohm}, {de O{\~n}a Wilhelmi}, {Opitz},
  {Ostrowski}, {Oya}, {Panter}, {Paz Arribas}, {Pedaletti}, {Pelletier},
  {Petrucci}, {Pita}, {P{\"u}hlhofer}, {Punch}, {Quirrenbach}, {Raue},
  {Rayner}, {Reimer}, {Reimer}, {Renaud}, {de los Reyes}, {Rieger}, {Ripken},
  {Rob}, {Rosier-Lees}, {Rowell}, {Rudak}, {Rulten}, {Ruppel}, {Sahakian},
  {Sanchez}, {Santangelo}, {Schlickeiser}, {Sch{\"o}ck}, {Schulz}, {Schwanke},
  {Schwarzburg}, {Schwemmer}, {Sheidaei}, {Skilton}, {Sol}, {Spengler},
  {Stawarz}, {Steenkamp}, {Stegmann}, {Stinzing}, {Stycz}, {Sushch}, {Szostek},
  {Tavernet}, {Terrier}, {Tluczykont}, {Valerius}, {van Eldik}, {Vasileiadis},
  {Venter}, {Vialle}, {Viana}, {Vincent}, {V{\"o}lk}, {Volpe}, {Vorobiov},
  {Vorster}, {Wagner}, {Ward}, {White}, {Wierzcholska}, {Zacharias}, {Zajczyk},
  {Zdziarski}, {Zech}, {Zechlin}, {H.~E.~S.~S. Collaboration}, {Aleksi{\'c}},
  {Antonelli}, {Antoranz}, {Backes}, {Barrio}, {Bastieri}, {Becerra
  Gonz{\'a}lez}, {Bednarek}, {Berdyugin}, {Berger}, {Bernardini}, {Biland},
  {Blanch}, {Bock}, {Boller}, {Bonnoli}, {Borla Tridon}, {Braun}, {Bretz},
  {Ca{\~n}ellas}, {Carmona}, {Carosi}, {Colin}, {Colombo}, {Contreras},
  {Cortina}, {Cossio}, {Covino}, {Dazzi}, {De Angelis}, {De Cea del Pozo}, {De
  Lotto}, {Delgado Mendez}, {Diago Ortega}, {Doert}, {Dom{\'\i}nguez}, {Dominis
  Prester}, {Dorner}, {Doro}, {Elsaesser}, {Ferenc}, {Fonseca}, {Font},
  {Fruck}, {Garc{\'\i}a L{\'o}pez}, {Garczarczyk}, {Garrido}, {Giavitto},
  {Godinovi{\'c}}, {Hadasch}, {H{\"a}fner}, {Herrero}, {Hildebrand},
  {H{\"o}hne-M{\"o}nch}, {Hose}, {Hrupec}, {Huber}, {Jogler}, {Klepser},
  {Kr{\"a}henb{\"u}hl}, {Krause}, {La Barbera}, {Lelas}, {Leonardo},
  {Lindfors}, {Lombardi}, {L{\'o}pez}, {Lorenz}, {Makariev}, {Maneva},
  {Mankuzhiyil}, {Mannheim}, {Maraschi}, {Mariotti}, {Mart{\'\i}nez}, {Mazin},
  {Meucci}, {Miranda}, {Mirzoyan}, {Miyamoto}, {Mold{\'o}n}, {Moralejo},
  {Munar}, {Nieto}, {Nilsson}, {Orito}, {Oya}, {Paneque}, {Paoletti}, {Pardo},
  {Paredes}, {Partini}, {Pasanen}, {Pauss}, {Perez-Torres}, {Persic},
  {Peruzzo}, {Pilia}, {Pochon}, {Prada}, {Prada Moroni}, {Prandini}, {Puljak},
  {Reichardt}, {Reinthal}, {Rhode}, {Rib{\'o}}, {Rico}, {R{\"u}gamer},
  {Saggion}, {Saito}, {Saito}, {Salvati}, {Satalecka}, {Scalzotto}, {Scapin},
  {Schultz}, {Schweizer}, {Shayduk}, {Shore}, {Sillanp{\"a}{\"a}}, {Sitarek},
  {Sobczynska}, {Spanier}, {Spiro}, {Stamerra}, {Steinke}, {Storz}, {Strah},
  {Suri{\'c}}, {Takalo}, {Takami}, {Tavecchio}, {Temnikov}, {Terzi{\'c}},
  {Tescaro}, {Teshima}, {Thom}, {Tibolla}, {Torres}, {Treves}, {Vankov},
  {Vogler}, {Wagner}, {Weitzel}, {Zabalza}, {Zandanel}, {Zanin}, {MAGIC
  Collaboration}, {Arlen}, {Aune}, {Beilicke}, {Benbow}, {Bouvier}, {Bradbury},
  {Buckley}, {Bugaev}, {Byrum}, {Cannon}, {Cesarini}, {Ciupik}, {Connolly},
  {Cui}, {Dickherber}, {Duke}, {Errando}, {Falcone}, {Finley}, {Finnegan},
  {Fortson}, {Furniss}, {Galante}, {Gall}, {Godambe}, {Griffin}, {Grube},
  {Gyuk}, {Hanna}, {Holder}, {Huan}, {Hui}, {Kaaret}, {Karlsson}, {Kertzman},
  {Khassen}, {Kieda}, {Krawczynski}, {Krennrich}, {Lang}, {LeBohec}, {Maier},
  {McArthur}, {McCann}, {Moriarty}, {Mukherjee}, {Nu{\~n}ez}, {Ong}, {Orr},
  {Otte}, {Park}, {Perkins}, {Pichel}, {Pohl}, {Prokoph}, {Ragan}, {Reyes},
  {Reynolds}, {Roache}, {Rose}, {Ruppel}, {Schroedter}, {Sembroski},
  {{\c{S}}ent{\"u}rk}, {Telezhinsky}, {Te{\v{s}}i{\'c}}, {Theiling},
  {Thibadeau}, {Varlotta}, {Vassiliev}, {Vivier}, {Wakely}, {Weekes},
  {Williams}, {Zitzer}, {VERITAS Collaboration}, {Barres de Almeida}, {Cara},
  {Casadio}, {Cheung}, {McConville}, {Davies}, {Doi}, {Giovannini},
  {Giroletti}, {Hada}, {Hardee}, {Harris}, {Junor}, {Kino}, {Lee}, {Ly},
  {Madrid}, {Massaro}, {Mundell}, {Nagai}, {Perlman}, {Steele}, {Walker}, and
  {Wood}]{2012ApJ...746..151A}
{Abramowski}, A.; {Acero}, F.; {Aharonian}, F.; {Akhperjanian}, A.G.; {Anton},
  G.; {Balzer}, A.; {Barnacka}, A.; {Barres de Almeida}, U.; {Becherini}, Y.;
  {Becker}, J.;  et~al.
\newblock {The 2010 Very High Energy {\ensuremath{\gamma}}-Ray Flare and 10
  Years of Multi-wavelength Observations of M 87}.
\newblock {\em \apj} {\bf 2012}, {\em 746},~151,
  \href{http://xxx.lanl.gov/abs/1111.5341}{{\normalfont
  [arXiv:astro-ph.CO/1111.5341]}}.
\newblock {\url{https://doi.org/10.1088/0004-637X/746/2/151}}.

\bibitem[{Biretta} \em{et~al.}(1991){Biretta}, {Stern}, and
  {Harris}]{1991AJ....101.1632B}
{Biretta}, J.A.; {Stern}, C.P.; {Harris}, D.E.
\newblock {The Radio to X-ray Spectrum of the M87 Jet and Nucleus}.
\newblock {\em \aj} {\bf 1991}, {\em 101},~1632.
\newblock {\url{https://doi.org/10.1086/115793}}.

\bibitem[{Palmer} \em{et~al.}(1967){Palmer}, {Rowson}, {Anderson}, {Donaldson},
  and {Miley}]{1967Natur.213..789P}
{Palmer}, H.P.; {Rowson}, B.; {Anderson}, B.; {Donaldson}, W.; {Miley}, G.K.
\newblock {Radio Diameter Measurements with Interferometer Baselines of One
  Million and Two Million Wavelengths}.
\newblock {\em \nat} {\bf 1967}, {\em 213},~789--790.
\newblock {\url{https://doi.org/10.1038/213789a0}}.

\bibitem[{Reid} \em{et~al.}(1982){Reid}, {Schmitt}, {Owen}, {Booth},
  {Wilkinson}, {Shaffer}, {Johnston}, and {Hardee}]{1982ApJ...263..615R}
{Reid}, M.J.; {Schmitt}, J.H.M.M.; {Owen}, F.N.; {Booth}, R.S.; {Wilkinson},
  P.N.; {Shaffer}, D.B.; {Johnston}, K.J.; {Hardee}, P.E.
\newblock {VLBI observations of the nucleus and jet of M 87.}
\newblock {\em \apj} {\bf 1982}, {\em 263},~615--623.
\newblock {\url{https://doi.org/10.1086/160533}}.

\bibitem[{Kovalev} \em{et~al.}(2007){Kovalev}, {Lister}, {Homan}, and
  {Kellermann}]{2007ApJ...668L..27K}
{Kovalev}, Y.Y.; {Lister}, M.L.; {Homan}, D.C.; {Kellermann}, K.I.
\newblock {The Inner Jet of the Radio Galaxy M87}.
\newblock {\em \apjl} {\bf 2007}, {\em 668},~L27--L30,
  \href{http://xxx.lanl.gov/abs/0708.2695}{{\normalfont
  [arXiv:astro-ph/0708.2695]}}.
\newblock {\url{https://doi.org/10.1086/522603}}.

\bibitem[{Kuo} \em{et~al.}(2014){Kuo}, {Asada}, {Rao}, {Nakamura}, {Algaba},
  {Liu}, {Inoue}, {Koch}, {Ho}, {Matsushita}, {Pu}, {Akiyama}, {Nishioka}, and
  {Pradel}]{2014ApJ...783L..33K}
{Kuo}, C.Y.; {Asada}, K.; {Rao}, R.; {Nakamura}, M.; {Algaba}, J.C.; {Liu},
  H.B.; {Inoue}, M.; {Koch}, P.M.; {Ho}, P.T.P.; {Matsushita}, S.;  et~al.
\newblock {Measuring Mass Accretion Rate onto the Supermassive Black Hole in
  M87 Using Faraday Rotation Measure with the Submillimeter Array}.
\newblock {\em \apjl} {\bf 2014}, {\em 783},~L33,
  \href{http://xxx.lanl.gov/abs/1402.5238}{{\normalfont
  [arXiv:astro-ph.GA/1402.5238]}}.
\newblock {\url{https://doi.org/10.1088/2041-8205/783/2/L33}}.

\bibitem[{Kino} \em{et~al.}(2015){Kino}, {Takahara}, {Hada}, {Akiyama},
  {Nagai}, and {Sohn}]{2015ApJ...803...30K}
{Kino}, M.; {Takahara}, F.; {Hada}, K.; {Akiyama}, K.; {Nagai}, H.; {Sohn},
  B.W.
\newblock {Magnetization Degree at the Jet Base of M87 Derived from the Event
  Horizon Telescope Data: Testing the Magnetically Driven Jet Paradigm}.
\newblock {\em \apj} {\bf 2015}, {\em 803},~30,
  \href{http://xxx.lanl.gov/abs/1502.03900}{{\normalfont
  [arXiv:astro-ph.HE/1502.03900]}}.
\newblock {\url{https://doi.org/10.1088/0004-637X/803/1/30}}.

\bibitem[{Mertens} \em{et~al.}(2016){Mertens}, {Lobanov}, {Walker}, and
  {Hardee}]{2016A&A...595A..54M}
{Mertens}, F.; {Lobanov}, A.P.; {Walker}, R.C.; {Hardee}, P.E.
\newblock {Kinematics of the jet in M 87 on scales of 100-1000 Schwarzschild
  radii}.
\newblock {\em \aap} {\bf 2016}, {\em 595},~A54,
  \href{http://xxx.lanl.gov/abs/1608.05063}{{\normalfont
  [arXiv:astro-ph.HE/1608.05063]}}.
\newblock {\url{https://doi.org/10.1051/0004-6361/201628829}}.

\bibitem[{Walker} \em{et~al.}(2018){Walker}, {Hardee}, {Davies}, {Ly}, and
  {Junor}]{2018ApJ...855..128W}
{Walker}, R.C.; {Hardee}, P.E.; {Davies}, F.B.; {Ly}, C.; {Junor}, W.
\newblock {The Structure and Dynamics of the Subparsec Jet in M87 Based on 50
  VLBA Observations over 17 Years at 43 GHz}.
\newblock {\em \apj} {\bf 2018}, {\em 855},~128,
  \href{http://xxx.lanl.gov/abs/1802.06166}{{\normalfont
  [arXiv:astro-ph.HE/1802.06166]}}.
\newblock {\url{https://doi.org/10.3847/1538-4357/aaafcc}}.

\bibitem[{Kim} \em{et~al.}(2018){Kim}, {Krichbaum}, {Lu}, {Ros}, {Bach},
  {Bremer}, {de Vicente}, {Lindqvist}, and {Zensus}]{2018A&A...616A.188K}
{Kim}, J.Y.; {Krichbaum}, T.P.; {Lu}, R.S.; {Ros}, E.; {Bach}, U.; {Bremer},
  M.; {de Vicente}, P.; {Lindqvist}, M.; {Zensus}, J.A.
\newblock {The limb-brightened jet of M87 down to the 7 Schwarzschild radii
  scale}.
\newblock {\em \aap} {\bf 2018}, {\em 616},~A188,
  \href{http://xxx.lanl.gov/abs/1805.02478}{{\normalfont
  [arXiv:astro-ph.GA/1805.02478]}}.
\newblock {\url{https://doi.org/10.1051/0004-6361/201832921}}.

\bibitem[{Chael} \em{et~al.}(2019){Chael}, {Narayan}, and
  {Johnson}]{2019MNRAS.486.2873C}
{Chael}, A.; {Narayan}, R.; {Johnson}, M.D.
\newblock {Two-temperature, Magnetically Arrested Disc simulations of the jet
  from the supermassive black hole in M87}.
\newblock {\em \mnras} {\bf 2019}, {\em 486},~2873--2895,
  \href{http://xxx.lanl.gov/abs/1810.01983}{{\normalfont
  [arXiv:astro-ph.HE/1810.01983]}}.
\newblock {\url{https://doi.org/10.1093/mnras/stz988}}.

\bibitem[{Meisenheimer} \em{et~al.}(1996){Meisenheimer}, {Roeser}, and
  {Schloetelburg}]{1996A&A...307...61M}
{Meisenheimer}, K.; {Roeser}, H.J.; {Schloetelburg}, M.
\newblock {The synchrotron spectrum of the jet in M87.}
\newblock {\em \aap} {\bf 1996}, {\em 307},~61.

\bibitem[{Kim} \em{et~al.}(2020){Kim}, {Krichbaum}, {Broderick}, {Wielgus},
  {Blackburn}, {G{\'o}mez}, {Johnson}, {Bouman}, {Chael}, {Akiyama}, {Jorstad},
  {Marscher}, {Issaoun}, {Janssen}, {Chan}, {Savolainen}, {Pesce}, {{\"O}zel},
  and {Event Horizon Telescope Collaboration}]{2020A&A...640A..69K}
{Kim}, J.Y.; {Krichbaum}, T.P.; {Broderick}, A.E.; {Wielgus}, M.; {Blackburn},
  L.; {G{\'o}mez}, J.L.; {Johnson}, M.D.; {Bouman}, K.L.; {Chael}, A.;
  {Akiyama}, K.;  et~al.
\newblock {Event Horizon Telescope imaging of the archetypal blazar 3C 279 at
  an extreme 20 microarcsecond resolution}.
\newblock {\em \aap} {\bf 2020}, {\em 640},~A69.
\newblock {\url{https://doi.org/10.1051/0004-6361/202037493}}.

\bibitem[{Wardle} \em{et~al.}(1998){Wardle}, {Homan}, {Ojha}, and
  {Roberts}]{1998Natur.395..457W}
{Wardle}, J.F.C.; {Homan}, D.C.; {Ojha}, R.; {Roberts}, D.H.
\newblock {Electron-positron jets associated with the quasar 3C279}.
\newblock {\em \nat} {\bf 1998}, {\em 395},~457--461.
\newblock {\url{https://doi.org/10.1038/26675}}.

\bibitem[{Hirotani} \em{et~al.}(2000){Hirotani}, {Iguchi}, {Kimura}, and
  {Wajima}]{2000ApJ...545..100H}
{Hirotani}, K.; {Iguchi}, S.; {Kimura}, M.; {Wajima}, K.
\newblock {Pair Plasma Dominance in the Parsec-Scale Relativistic Jet of 3C
  345}.
\newblock {\em \apj} {\bf 2000}, {\em 545},~100--106,
  \href{http://xxx.lanl.gov/abs/astro-ph/0005394}{{\normalfont
  [arXiv:astro-ph/astro-ph/0005394]}}.
\newblock {\url{https://doi.org/10.1086/317769}}.

\bibitem[{Marscher} \em{et~al.}(2007){Marscher}, {Jorstad}, {G{\'o}mez},
  {McHardy}, {Krichbaum}, and {Agudo}]{2007ApJ...665..232M}
{Marscher}, A.P.; {Jorstad}, S.G.; {G{\'o}mez}, J.L.; {McHardy}, I.M.;
  {Krichbaum}, T.P.; {Agudo}, I.
\newblock {Search for Electron-Positron Annihilation Radiation from the Jet in
  3C 120}.
\newblock {\em \apj} {\bf 2007}, {\em 665},~232--236.
\newblock {\url{https://doi.org/10.1086/519481}}.

\bibitem[{Celotti} and {Fabian}(1993)]{1993MNRAS.264..228C}
{Celotti}, A.; {Fabian}, A.C.
\newblock {The kinetic power and luminosity of parsec-scale radio jets - an
  argument for heavy jets.}
\newblock {\em \mnras} {\bf 1993}, {\em 264},~228--236.
\newblock {\url{https://doi.org/10.1093/mnras/264.1.228}}.

\bibitem[{Broderick} and {Tchekhovskoy}(2015)]{2015ApJ...809...97B}
{Broderick}, A.E.; {Tchekhovskoy}, A.
\newblock {Horizon-scale Lepton Acceleration in Jets: Explaining the Compact
  Radio Emission in M87}.
\newblock {\em \apj} {\bf 2015}, {\em 809},~97,
  \href{http://xxx.lanl.gov/abs/1506.04754}{{\normalfont
  [arXiv:astro-ph.HE/1506.04754]}}.
\newblock {\url{https://doi.org/10.1088/0004-637X/809/1/97}}.

\bibitem[{Breit} and {Wheeler}(1934)]{1934PhRv...46.1087B}
{Breit}, G.; {Wheeler}, J.A.
\newblock {Collision of Two Light Quanta}.
\newblock {\em Physical Review} {\bf 1934}, {\em 46},~1087--1091.
\newblock {\url{https://doi.org/10.1103/PhysRev.46.1087}}.

\bibitem[{Beskin} \em{et~al.}(1992){Beskin}, {Istomin}, and
  {Parev}]{1992SvA....36..642B}
{Beskin}, V.S.; {Istomin}, Y.N.; {Parev}, V.I.
\newblock {Filling the Magnetosphere of a Supermassive Black-Hole with Plasma}.
\newblock {\em \sovast} {\bf 1992}, {\em 36},~642.

\bibitem[{Hirotani} and {Okamoto}(1998)]{1998ApJ...497..563H}
{Hirotani}, K.; {Okamoto}, I.
\newblock {Pair Plasma Production in a Force-free Magnetosphere around a
  Supermassive Black Hole}.
\newblock {\em \apj} {\bf 1998}, {\em 497},~563--572.
\newblock {\url{https://doi.org/10.1086/305479}}.

\bibitem[{Ford} \em{et~al.}(2018){Ford}, {Keenan}, and
  {Medvedev}]{2018PhRvD..98f3016F}
{Ford}, A.L.; {Keenan}, B.D.; {Medvedev}, M.V.
\newblock {Electron-positron cascade in magnetospheres of spinning black
  holes}.
\newblock {\em \prd} {\bf 2018}, {\em 98},~063016,
  \href{http://xxx.lanl.gov/abs/1706.00542}{{\normalfont
  [arXiv:astro-ph.HE/1706.00542]}}.
\newblock {\url{https://doi.org/10.1103/PhysRevD.98.063016}}.

\bibitem[{Levinson} and {Cerutti}(2018)]{2018A&A...616A.184L}
{Levinson}, A.; {Cerutti}, B.
\newblock {Particle-in-cell simulations of pair discharges in a starved
  magnetosphere of a Kerr black hole}.
\newblock {\em \aap} {\bf 2018}, {\em 616},~A184,
  \href{http://xxx.lanl.gov/abs/1803.04427}{{\normalfont
  [arXiv:astro-ph.HE/1803.04427]}}.
\newblock {\url{https://doi.org/10.1051/0004-6361/201832915}}.

\bibitem[{Chen} \em{et~al.}(2018){Chen}, {Yuan}, and
  {Yang}]{2018ApJ...863L..31C}
{Chen}, A.Y.; {Yuan}, Y.; {Yang}, H.
\newblock {Physics of Pair Producing Gaps in Black Hole Magnetospheres}.
\newblock {\em \apjl} {\bf 2018}, {\em 863},~L31,
  \href{http://xxx.lanl.gov/abs/1805.11039}{{\normalfont
  [arXiv:astro-ph.HE/1805.11039]}}.
\newblock {\url{https://doi.org/10.3847/2041-8213/aad8ab}}.

\bibitem[{Parfrey} \em{et~al.}(2019){Parfrey}, {Philippov}, and
  {Cerutti}]{2019PhRvL.122c5101P}
{Parfrey}, K.; {Philippov}, A.; {Cerutti}, B.
\newblock {First-Principles Plasma Simulations of Black-Hole Jet Launching}.
\newblock {\em \prl} {\bf 2019}, {\em 122},~035101,
  \href{http://xxx.lanl.gov/abs/1810.03613}{{\normalfont
  [arXiv:astro-ph.HE/1810.03613]}}.
\newblock {\url{https://doi.org/10.1103/PhysRevLett.122.035101}}.

\bibitem[{Mo{\'s}cibrodzka} \em{et~al.}(2011){Mo{\'s}cibrodzka}, {Gammie},
  {Dolence}, and {Shiokawa}]{2011ApJ...735....9M}
{Mo{\'s}cibrodzka}, M.; {Gammie}, C.F.; {Dolence}, J.C.; {Shiokawa}, H.
\newblock {Pair Production in Low-luminosity Galactic Nuclei}.
\newblock {\em \apj} {\bf 2011}, {\em 735},~9,
  \href{http://xxx.lanl.gov/abs/1104.2042}{{\normalfont
  [arXiv:astro-ph.HE/1104.2042]}}.
\newblock {\url{https://doi.org/10.1088/0004-637X/735/1/9}}.

\bibitem[{Wong} \em{et~al.}(2021){Wong}, {Ryan}, and
  {Gammie}]{2021ApJ...907...73W}
{Wong}, G.N.; {Ryan}, B.R.; {Gammie}, C.F.
\newblock {Pair Drizzle around Sub-Eddington Supermassive Black Holes}.
\newblock {\em \apj} {\bf 2021}, {\em 907},~73,
  \href{http://xxx.lanl.gov/abs/2012.04658}{{\normalfont
  [arXiv:astro-ph.HE/2012.04658]}}.
\newblock {\url{https://doi.org/10.3847/1538-4357/abd0f9}}.

\bibitem[{Ryan} \em{et~al.}(2015){Ryan}, {Dolence}, and
  {Gammie}]{2015ApJ...807...31R}
{Ryan}, B.R.; {Dolence}, J.C.; {Gammie}, C.F.
\newblock {bhlight: General Relativistic Radiation Magnetohydrodynamics with
  Monte Carlo Transport}.
\newblock {\em \apj} {\bf 2015}, {\em 807},~31,
  \href{http://xxx.lanl.gov/abs/1505.05119}{{\normalfont
  [arXiv:astro-ph.HE/1505.05119]}}.
\newblock {\url{https://doi.org/10.1088/0004-637X/807/1/31}}.

\bibitem[{Ryan} \em{et~al.}(2017){Ryan}, {Ressler}, {Dolence}, {Tchekhovskoy},
  {Gammie}, and {Quataert}]{2017ApJ...844L..24R}
{Ryan}, B.R.; {Ressler}, S.M.; {Dolence}, J.C.; {Tchekhovskoy}, A.; {Gammie},
  C.; {Quataert}, E.
\newblock {The Radiative Efficiency and Spectra of Slowly Accreting Black Holes
  from Two-temperature GRRMHD Simulations}.
\newblock {\em \apjl} {\bf 2017}, {\em 844},~L24,
  \href{http://xxx.lanl.gov/abs/1707.04238}{{\normalfont
  [arXiv:astro-ph.HE/1707.04238]}}.
\newblock {\url{https://doi.org/10.3847/2041-8213/aa8034}}.

\bibitem[{Ryan} \em{et~al.}(2018){Ryan}, {Ressler}, {Dolence}, {Gammie}, and
  {Quataert}]{2018ApJ...864..126R}
{Ryan}, B.R.; {Ressler}, S.M.; {Dolence}, J.C.; {Gammie}, C.; {Quataert}, E.
\newblock {Two-temperature GRRMHD Simulations of M87}.
\newblock {\em \apj} {\bf 2018}, {\em 864},~126,
  \href{http://xxx.lanl.gov/abs/1808.01958}{{\normalfont
  [arXiv:astro-ph.HE/1808.01958]}}.
\newblock {\url{https://doi.org/10.3847/1538-4357/aad73a}}.

\bibitem[{Ryan} \em{et~al.}(2019){Ryan}, {Dolence}, {Gammie}, {Ressler}, and
  {Miller}]{2019ascl.soft09007R}
{Ryan}, B.R.; {Dolence}, J.C.; {Gammie}, C.F.; {Ressler}, S.M.; {Miller}, J.
\newblock {EBHLIGHT: General relativistic radiation magnetohydrodynamics with
  Monte Carlo transport}.
\newblock Astrophysics Source Code Library, record ascl:1909.007,  2019,
  \href{http://xxx.lanl.gov/abs/1909.007}{{\normalfont [1909.007]}}.

\bibitem[{Blandford} and {K{\"o}nigl}(1979)]{Blandford&Konigl1979}
{Blandford}, R.D.; {K{\"o}nigl}, A.
\newblock {Relativistic jets as compact radio sources.}
\newblock {\em \apj} {\bf 1979}, {\em 232},~34--48.
\newblock {\url{https://doi.org/10.1086/157262}}.

\bibitem[{Wardle} \em{et~al.}(1998){Wardle}, {Homan}, {Ojha}, and
  {Roberts}]{Wardle+1998}
{Wardle}, J.F.C.; {Homan}, D.C.; {Ojha}, R.; {Roberts}, D.H.
\newblock {Electron-positron jets associated with the quasar 3C279}.
\newblock {\em \nat} {\bf 1998}, {\em 395},~457--461.
\newblock {\url{https://doi.org/10.1038/26675}}.

\bibitem[{Wardle} and {Homan}(2003)]{Wardle&Homan2003}
{Wardle}, J.F.C.; {Homan}, D.C.
\newblock {Theoretical Models for Producing Circularly Polarized Radiation in
  Extragalactic Radio Sources}.
\newblock {\em \apss} {\bf 2003}, {\em 288},~143--153,
  \href{http://xxx.lanl.gov/abs/astro-ph/0305136}{{\normalfont
  [arXiv:astro-ph/astro-ph/0305136]}}.
\newblock {\url{https://doi.org/10.1023/B:ASTR.0000005001.80514.0c}}.

\bibitem[{Dexter}(2016)]{Dexter2016}
{Dexter}, J.
\newblock {A public code for general relativistic, polarised radiative transfer
  around spinning black holes}.
\newblock {\em \mnras} {\bf 2016}, {\em 462},~115--136,
  \href{http://xxx.lanl.gov/abs/1602.03184}{{\normalfont
  [arXiv:astro-ph.HE/1602.03184]}}.
\newblock {\url{https://doi.org/10.1093/mnras/stw1526}}.

\bibitem[{Emami} \em{et~al.}(2021){Emami}, {Anantua}, {Chael}, and
  {Loeb}]{2021ApJ...923..272E}
{Emami}, R.; {Anantua}, R.; {Chael}, A.A.; {Loeb}, A.
\newblock {Positron Effects on Polarized Images and Spectra from Jet and
  Accretion Flow Models of M87* and Sgr A*}.
\newblock {\em \apj} {\bf 2021}, {\em 923},~272,
  \href{http://xxx.lanl.gov/abs/2101.05327}{{\normalfont
  [arXiv:astro-ph.HE/2101.05327]}}.
\newblock {\url{https://doi.org/10.3847/1538-4357/ac2950}}.

\bibitem[Anantua \em{et~al.}(2020)Anantua, Emami, Loeb, and
  Chael]{Anantua2020a}
Anantua, R.; Emami, R.; Loeb, A.; Chael, A.
\newblock {Determining the Composition of Relativistic Jets from Polarization
  Maps}.
\newblock {\em Astrophys. J.} {\bf 2020}, {\em 896},~30,
  \href{http://xxx.lanl.gov/abs/1909.09230}{{\normalfont
  [arXiv:astro-ph.HE/1909.09230]}}.
\newblock {\url{https://doi.org/10.3847/1538-4357/ab9103}}.

\bibitem[Blandford and Anantua(2017)]{Blandford2017}
Blandford, R.; Anantua, R.
\newblock The Future of Black Hole Astrophysics in the {LIGO}-{VIRGO}-{LPF}
  Era.
\newblock {\em Journal of Physics: Conference Series} {\bf 2017}, {\em
  840},~012023.
\newblock {\url{https://doi.org/10.1088/1742-6596/840/1/012023}}.

\bibitem[McKinney \em{et~al.}(2012)McKinney, Tchekhovskoy, and
  Blandford]{McKinney2012}
McKinney, J.C.; Tchekhovskoy, A.; Blandford, R.D.
\newblock General relativistic magnetohydrodynamic simulations of magnetically
  choked accretion flows around black holes.
\newblock {\em Monthly Notices of the Royal Astronomical Society} {\bf 2012},
  {\em 423},~3083–3117.
\newblock {\url{https://doi.org/10.1111/j.1365-2966.2012.21074.x}}.

\bibitem[{Event Horizon Telescope Collaboration} \em{et~al.}(2021){Event
  Horizon Telescope Collaboration}, {Akiyama}, {Algaba}, {Alberdi}, {Alef},
  {Anantua}, {Asada}, {Azulay}, {Baczko}, {Ball}, {Balokovi{\'c}}, {Barrett},
  {Benson}, {Bintley}, {Blackburn}, {Blundell}, {Boland}, {Bouman}, {Bower},
  {Boyce}, {Bremer}, {Brinkerink}, {Brissenden}, {Britzen}, {Broderick},
  {Broguiere}, {Bronzwaer}, {Byun}, {Carlstrom}, {Chael}, {Chan}, {Chatterjee},
  {Chatterjee}, {Chen}, {Chen}, {Chesler}, {Cho}, {Christian}, {Conway},
  {Cordes}, {Crawford}, {Crew}, {Cruz-Osorio}, {Cui}, {Davelaar}, {De
  Laurentis}, {Deane}, {Dempsey}, {Desvignes}, {Dexter}, {Doeleman}, {Eatough},
  {Falcke}, {Farah}, {Fish}, {Fomalont}, {Ford}, {Fraga-Encinas}, {Friberg},
  {Fromm}, {Fuentes}, {Galison}, {Gammie}, {Garc{\'\i}a}, {Gelles}, {Gentaz},
  {Georgiev}, {Goddi}, {Gold}, {G{\'o}mez}, {G{\'o}mez-Ruiz}, {Gu}, {Gurwell},
  {Hada}, {Haggard}, {Hecht}, {Hesper}, {Himwich}, {Ho}, {Ho}, {Honma},
  {Huang}, {Huang}, {Hughes}, {Ikeda}, {Inoue}, {Issaoun}, {James}, {Jannuzi},
  {Janssen}, {Jeter}, {Jiang}, {Jimenez-Rosales}, {Johnson}, {Jorstad}, {Jung},
  {Karami}, {Karuppusamy}, {Kawashima}, {Keating}, {Kettenis}, {Kim}, {Kim},
  {Kim}, {Kim}, {Kino}, {Koay}, {Kofuji}, {Koch}, {Koyama}, {Kramer}, {Kramer},
  {Krichbaum}, {Kuo}, {Lauer}, {Lee}, {Levis}, {Li}, {Li}, {Lindqvist}, {Lico},
  {Lindahl}, {Liu}, {Liu}, {Liuzzo}, {Lo}, {Lobanov}, {Loinard}, {Lonsdale},
  {Lu}, {MacDonald}, {Mao}, {Marchili}, {Markoff}, {Marrone}, {Marscher},
  {Mart{\'\i}-Vidal}, {Matsushita}, {Matthews}, {Medeiros}, {Menten}, {Mizuno},
  {Mizuno}, {Moran}, {Moriyama}, {Moscibrodzka}, {M{\"u}ller}, {Musoke}, {Mus
  Mej{\'\i}as}, {Michalik}, {Nadolski}, {Nagai}, {Nagar}, {Nakamura},
  {Narayan}, {Narayanan}, {Natarajan}, {Nathanail}, {Neilsen}, {Neri}, {Ni},
  {Noutsos}, {Nowak}, {Okino}, {Olivares}, {Ortiz-Le{\'o}n}, {Oyama},
  {{\"O}zel}, {Palumbo}, {Park}, {Patel}, {Pen}, {Pesce}, {Pi{\'e}tu},
  {Plambeck}, {PopStefanija}, {Porth}, {P{\"o}tzl}, {Prather},
  {Preciado-L{\'o}pez}, {Psaltis}, {Pu}, {Ramakrishnan}, {Rao}, {Rawlings},
  {Raymond}, {Rezzolla}, {Ricarte}, {Ripperda}, {Roelofs}, {Rogers}, {Ros},
  {Rose}, {Roshanineshat}, {Rottmann}, {Roy}, {Ruszczyk}, {Rygl},
  {S{\'a}nchez}, {S{\'a}nchez-Arguelles}, {Sasada}, {Savolainen}, {Schloerb},
  {Schuster}, {Shao}, {Shen}, {Small}, {Sohn}, {SooHoo}, {Sun}, {Tazaki},
  {Tetarenko}, {Tiede}, {Tilanus}, {Titus}, {Toma}, {Torne}, {Trent},
  {Traianou}, {Trippe}, {van Bemmel}, {van Langevelde}, {van Rossum}, {Wagner},
  {Ward-Thompson}, {Wardle}, {Weintroub}, {Wex}, {Wharton}, {Wielgus}, {Wong},
  {Wu}, {Yoon}, {Young}, {Young}, {Younsi}, {Yuan}, {Yuan}, {Zensus}, {Zhao},
  and {Zhao}]{2021ApJ...910L..13E}
{Event Horizon Telescope Collaboration}.; {Akiyama}, K.; {Algaba}, J.C.;
  {Alberdi}, A.; {Alef}, W.; {Anantua}, R.; {Asada}, K.; {Azulay}, R.;
  {Baczko}, A.K.; {Ball}, D.;  et~al.
\newblock {First M87 Event Horizon Telescope Results. VIII. Magnetic Field
  Structure near The Event Horizon}.
\newblock {\em \apjl} {\bf 2021}, {\em 910},~L13,
  \href{http://xxx.lanl.gov/abs/2105.01173}{{\normalfont
  [arXiv:astro-ph.HE/2105.01173]}}.
\newblock {\url{https://doi.org/10.3847/2041-8213/abe4de}}.

\bibitem[Doeleman \em{et~al.}(2008)Doeleman, Weintroub, Rogers, Plambeck,
  Freund, Tilanus, Friberg, Ziurys, Moran, Corey, and et~al.]{Doeleman2008}
Doeleman, S.S.; Weintroub, J.; Rogers, A.E.E.; Plambeck, R.; Freund, R.;
  Tilanus, R.P.J.; Friberg, P.; Ziurys, L.M.; Moran, J.M.; Corey, B.;  et~al.
\newblock Event-horizon-scale structure in the supermassive black hole
  candidate at the Galactic Centre.
\newblock {\em Nature} {\bf 2008}, {\em 455},~78–80.
\newblock {\url{https://doi.org/10.1038/nature07245}}.

\bibitem[{Doeleman} \em{et~al.}(2012){Doeleman}, {Fish}, {Schenck}, {Beaudoin},
  {Blundell}, {Bower}, {Broderick}, {Chamberlin}, {Freund}, {Friberg},
  {Gurwell}, {Ho}, {Honma}, {Inoue}, {Krichbaum}, {Lamb}, {Loeb}, {Lonsdale},
  {Marrone}, {Moran}, {Oyama}, {Plambeck}, {Primiani}, {Rogers}, {Smythe},
  {SooHoo}, {Strittmatter}, {Tilanus}, {Titus}, {Weintroub}, {Wright}, {Young},
  and {Ziurys}]{Doeleman2012}
{Doeleman}, S.S.; {Fish}, V.L.; {Schenck}, D.E.; {Beaudoin}, C.; {Blundell},
  R.; {Bower}, G.C.; {Broderick}, A.E.; {Chamberlin}, R.; {Freund}, R.;
  {Friberg}, P.;  et~al.
\newblock {Jet-Launching Structure Resolved Near the Supermassive Black Hole in
  M87}.
\newblock {\em Science} {\bf 2012}, {\em 338},~355,
  \href{http://xxx.lanl.gov/abs/1210.6132}{{\normalfont
  [arXiv:astro-ph.HE/1210.6132]}}.
\newblock {\url{https://doi.org/10.1126/science.1224768}}.

\bibitem[{Prieto} \em{et~al.}(2016){Prieto}, {Fern{\'a}ndez-Ontiveros},
  {Markoff}, {Espada}, and {Gonz{\'a}lez-Mart{\'\i}n}]{Prieto2016}
{Prieto}, M.A.; {Fern{\'a}ndez-Ontiveros}, J.A.; {Markoff}, S.; {Espada}, D.;
  {Gonz{\'a}lez-Mart{\'\i}n}, O.
\newblock {The central parsecs of M87: jet emission and an elusive accretion
  disc}.
\newblock {\em \mnras} {\bf 2016}, {\em 457},~3801--3816,
  \href{http://xxx.lanl.gov/abs/1508.02302}{{\normalfont
  [arXiv:astro-ph.GA/1508.02302]}}.
\newblock {\url{https://doi.org/10.1093/mnras/stw166}}.

\bibitem[{Prather} \em{et~al.}(2021){Prather}, {Wong}, {Dhruv}, {Ryan},
  {Dolence}, {Ressler}, and {Gammie}]{2021JOSS....6.3336P}
{Prather}, B.; {Wong}, G.; {Dhruv}, V.; {Ryan}, B.; {Dolence}, J.; {Ressler},
  S.; {Gammie}, C.
\newblock {iharm3D: Vectorized General Relativistic Magnetohydrodynamics}.
\newblock {\em The Journal of Open Source Software} {\bf 2021}, {\em 6},~3336,
  \href{http://xxx.lanl.gov/abs/2110.10191}{{\normalfont
  [arXiv:astro-ph.HE/2110.10191]}}.
\newblock {\url{https://doi.org/10.21105/joss.03336}}.

\bibitem[{Mo{\'s}cibrodzka} and {Gammie}(2018)]{Moscibrodzka&Gammie2018}
{Mo{\'s}cibrodzka}, M.; {Gammie}, C.F.
\newblock {IPOLE - semi-analytic scheme for relativistic polarized radiative
  transport}.
\newblock {\em \mnras} {\bf 2018}, {\em 475},~43--54,
  \href{http://xxx.lanl.gov/abs/1712.03057}{{\normalfont
  [arXiv:astro-ph.HE/1712.03057]}}.
\newblock {\url{https://doi.org/10.1093/mnras/stx3162}}.

\bibitem[{Ricarte} \em{et~al.}(2021){Ricarte}, {Qiu}, and
  {Narayan}]{2021MNRAS.505..523R}
{Ricarte}, A.; {Qiu}, R.; {Narayan}, R.
\newblock {Black hole magnetic fields and their imprint on circular
  polarization images}.
\newblock {\em \mnras} {\bf 2021}, {\em 505},~523--539,
  \href{http://xxx.lanl.gov/abs/2104.11301}{{\normalfont
  [arXiv:astro-ph.HE/2104.11301]}}.
\newblock {\url{https://doi.org/10.1093/mnras/stab1289}}.

\end{thebibliography}
 
\end{document}